\documentclass[superscriptaddress,aps,10pt,showkeys,notitlepage,pre]{revtex4-1}
\usepackage{times}
\usepackage[colorlinks,bookmarksopen,bookmarksnumbered,linkcolor=cyan,citecolor=cyan,urlcolor=blue,breaklinks=true]{hyperref}
\usepackage{bm}
\usepackage[english]{babel}
\usepackage{amssymb}
\usepackage{amsmath}
\usepackage{amsthm}
\usepackage{epsfig}
\usepackage{float}
\usepackage{array,multirow,makecell}
\usepackage{url}
\usepackage{color}
\usepackage{setspace}
\usepackage{anyfontsize} 

\protect
\protect
\newcommand{\etaref}{\eta^{\rm{ref}}}

\newcommand{\aref}{a^{\rm ref}}
\newcommand{\R}{\mathbb{R}}

\newcommand{\lt}{\left}
\newcommand{\rt}{\right}
\newcommand{\dx}{|\partial_x|}

\newcommand{\cnum}{c_{\bm{\eta}}}

\newcommand{\ee}{{\rm e}}
\newcommand{\dd}{{\rm d}}

\newcommand{\etal}{\eta_{\rm l}}

\newcommand{\etar}{\eta_{\rm r}}

\newcommand{\cW}{c_\eta}
\newcommand{\injec}{\mathcal{I}}
\newcommand{\projec}{\mathcal{P}}

\newcommand{\bigO}{{\rm O}}

\newcommand{\Fd}{\mathcal{F}_d}

\newcommand{\KK}{\mathcal{K}}

\newcommand{\dxnum}{\lt|D_x\rt|}
\newcommand{\pxnum}{D_x}

\begin{document}
\title{Fourier-based numerical approximation of the Weertman equation for moving dislocations}

\author{Marc \surname{Josien}}
\email{marc.josien@enpc.fr}
\affiliation{\'Ecole des Ponts and INRIA, 6 et 8 avenue Blaise Pascal, 77455 Marne-La-Vall\'ee Cedex 2, France.}

\author{Yves-Patrick \surname{Pellegrini}}
\email{yves-patrick.pellegrini@cea.fr}
\affiliation{CEA, DAM, DIF, F-91297 Arpajon, France.}

\author{Fr\'ed\'eric \surname{Legoll}}
\email{frederic.legoll@enpc.fr}
\affiliation{\'Ecole des Ponts and INRIA, 6 et 8 avenue Blaise Pascal, 77455 Marne-La-Vall\'ee Cedex 2, France.}

\author{Claude \surname{Le Bris}}
\email{lebris@cermics.enpc.fr}
\affiliation{\'Ecole des Ponts and INRIA, 6 et 8 avenue Blaise Pascal, 77455 Marne-La-Vall\'ee Cedex 2, France.}

\begin{abstract}
This work discusses the numerical approximation of a nonlinear reaction-advection-diffusion equation, which is a dimensionless form of the Weertman equation. This equation models steadily-moving dislocations in materials science. It reduces to the celebrated Peierls-Nabarro equation when its advection term is set to zero. The approach rests on considering a time-dependent formulation, which admits the equation under study as its long-time limit. Introducing a Preconditioned Collocation Scheme based on Fourier transforms, the iterative numerical method presented solves the time-dependent problem, delivering at convergence the desired numerical solution to the Weertman equation. Although it rests on an \emph{explicit} time-evolution scheme, the method allows for large time steps, and captures the solution in a robust manner. Numerical results illustrate the efficiency of the approach for several types of nonlinearities.
\end{abstract}


\date{\today}

\maketitle

\section{Introduction}
This article addresses the numerical approximation of the following nonlinear integrodifferential equation, with Cauchy-type singular kernel:
\begin{equation}
\label{Wr}
    \lt\{
	\begin{aligned}
	&-\lt|\partial_x \right| \eta(x)+ \cW\,\partial_x \eta(x) = F_\sigma'(\eta(x))
    \quad\text{for}\quad x\in\mathbb{R},\\
	    &\eta(-\infty)= \eta_{\rm l}\quad\text{and}\quad\eta(+\infty)=\eta_{\rm r},
	\end{aligned}
	\right.
\end{equation}
where both the real-valued function $\eta$ \emph{and} the scalar constant $\cW$ are the unknowns. The potential $F_\sigma$ is a nonlinear bistable function of $\eta$ with (at least) two local minima at values $\eta=\eta_{\rm l}$ and $\eta=\eta_{\rm r}$. The meaning of the subscript $\sigma$ is explained below. The operator $|\partial_x|$ is linear, and defined in terms of the Hilbert transform $\mathcal{H}$ \cite{KING09} as
\begin{align}
\label{eq:hilb}
|\partial_x|\eta(x)=\mathcal{H}(\partial_x\eta)(x)=\frac{1}{\pi}{\rm p.v.}\int_{-\infty}^{+\infty}\frac{\partial_x\eta(x')}{x-x'}\dd x'=\lim_{\epsilon \rightarrow 0}\frac{1}{\pi}\int_{|x-x'|>\epsilon}\frac{\partial_x\eta(x')}{x-x'}\dd x',
\end{align}
where ${\rm p.v.}$ denotes the principal value \cite{KANW04} at $x$. In the context of singular integral equations, the above kernel of the Hilbert transform is known as the Cauchy kernel. The operator $-|\partial_x|$, also denoted $-(-\Delta)^{1/2}$ by some authors, is diffusive \cite[p.\ 181]{Brezis}. Another useful representation of \eqref{eq:hilb} is (by integration by parts)
\begin{align}
\label{eq:hilb2}
|\partial_x|\eta(x)=-\frac{1}{\pi}\int_0^{+\infty}\frac{\eta(x+y)+\eta(x-y)-2\eta(x)}{y^2}\dd y.
\end{align}
Let the Fourier transform in the continuum (FT) of a function $f$ be defined at wavemode $k$ as
\begin{align}
  \mathcal{F}\{f\}(k)=\widehat{f}(k)=\int_{-\infty}^{+\infty} {\rm e}^{-{\rm i} k x}f(x)\,\dd x.
\end{align}
One has $\mathcal{F}\lt\{\text{p.v.}\,x^{-1}\rt\}(k)=-{\rm i}\,\pi\,{\rm{sgn}}(k)$ \cite[p.\ 1118]{Gradshteyn}, whence $\mathcal{F}\lt\{\,|\partial_x|\,\eta\rt\}(k)$ $=$ $|k|\widehat{\eta}(k)$. Thus, the non-local operator  $|\partial_x|$ is symmetric and positive.

\begin{figure}[ht]
    \begin{center}
	\includegraphics[width=6cm]{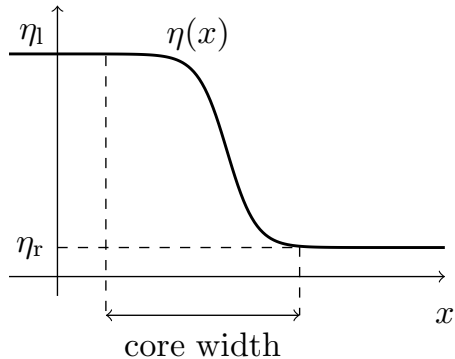}
    \end{center}
    \caption{\label{FigDisloc} Typical shape of $\eta(x)$ in Equation \eqref{Wr} when $F$ is a sinusoidal function.}
\end{figure}

Equation \eqref{Wr} is a dimensionless form of the Weertman equation \cite{WEER69a,WEER69b,Rosakis} (simply referred to as `the Weertman equation' in the following), which models straight dislocations traveling with steady velocity, thus generalizing the Peierls-Nabarro (PN) equation for static dislocations \cite{Nabarro46}:
\begin{equation}
\label{PN}
\lt\{
\begin{aligned}
    &-\lt|\partial_x \right| \eta(x)= F_\sigma'(\eta(x)) \quad\text{for}\quad x\in \mathbb{R},\\
	&\eta(-\infty)=\eta_{\rm l}\quad\text{and}\quad\eta(+\infty)=\eta_{\rm r}.
\end{aligned}
\right.
\end{equation}
Dislocations are linear defects in crystals, the motion of which is responsible for the plasticity of metals \cite{Hirth}. Dislocation lines have a non-vanishing sectional area, i.e., they possess a `core' of finite width.
The derivative $\partial_x\eta(x)$ (the so-called dislocation density) of the unknown function $\eta$ in \eqref{Wr} describes the shape function of a \emph{flat} finite-width dislocation on its glide plane, along the $x$-direction. The core is the region of space where $\partial_x\eta(x)$ develops peaks. From a physical standpoint, the function $\eta$ represents a local relative material displacement discontinuity between the upper and lower half-spaces surrounding the glide plane on which moves the dislocation line; see, e.g., \cite{Hirth} for details. From a broader perspective, the function $\eta$ can be understood as a moving phase-transformation front between the states $\eta_{\rm l}$ and $\eta_{\rm r}$ (Fig.\ \ref{FigDisloc}).

In \eqref{Wr} the term $|\partial_x| \eta$ accounts for the long-range elastic self-interactions that tend to spread the core. This repulsive interaction is counterbalanced by the nonlinear pull-back force $F_\sigma'(\eta)$, which binds together the upper and lower half-spaces, thus giving the dislocation core its finite width. Throughout this article, we consider that $F_\sigma'(\eta)$ includes a constant externally applied loading $\sigma$ (that is, $F_\sigma(\eta)=F(\eta)-\sigma\eta$ where $F(\eta)$ is an energy potential intrinsic to the material, tilted by the adjunction of a linear term $-\sigma\eta$). Moreover, the moving dislocation is subjected to various drag mechanisms encoded into the term $\cW\,\partial_x \eta$. As recalled below, the Weertman equation admits an analytical solution \cite{Rosakis} if $F(\eta)$ is assumed sinusoidal. For more realistic potentials, a numerical approach is required.

Since its inception, the original PN model has been enriched in various directions. For instance, it most often requires being generalized to vector-valued $\eta$ to be quantitatively predictive \cite{SydowHartfordWahnstrom,LuBulatovKioussis,LuBulatovKioussis2,DENO04,DENO07}. Also, the model has been extended to two dimensions of space, to study planar dislocation loops \cite{DENO04,DENO07,Xiang,zhu2015}. Quite generally, methods to compute the shape of static or moving cores encompass variational approaches and involve finite-element and/or phase-field-type implementations \cite{DENO04,DENO07,ZHAN15,Mianroodi}. Yet, in spite of such a wealth of enrichments of the PN model and its associated numerical methods of solution, the one-dimensional Weertman equation ---a comparatively simpler extension--- has not been investigated as thoroughly, while the specific problem of determining the allowed velocities of steadily-moving dislocations for general force laws $F_\sigma'$ remains an open question of major practical interest \cite{Rosakis}. For this reason, the present work focuses on solving the simplest, scalar, and one-dimensional case. A generalization to the vector case of the Weertman equation is the subject of ongoing work, and will be presented elsewhere \cite{YPP2}.

It must be emphasized that, in the dimensionless form \eqref{Wr} of the equation, $\cW$ is \emph{not} the physical velocity of the dislocation. The latter is deduced from $\cW$ in a post-processing step to be explained in \cite{YPP2} (it is a direct application of Equations (46), (47) or (48) of \cite{YPP}), which in the present scalar case is independent from the numerical task of solving the equation. Therefore the physical velocity is not further considered hereafter. \emph{Moreover, we stress that \eqref{Wr} applies only to \emph{subsonic} motion, for which the coefficient of $|\partial_x|$ does not vanish in the original Weertman equation}.

Numerical methods for solving integrodifferential equations such as \eqref{Wr} have been proposed by many authors. The method employed in \cite{Lejvcek} uses properties of the Hilbert transform to recast Equation \eqref{PN} into a form amenable to fixed-point approaches. In \cite{Kurtz} the authors consider a simpler version of \eqref{Wr} on a bounded domain, in which the nonlinear term $F_\sigma'(\eta(x))$ is replaced by some given $\eta$-independent function $g(x)$, and where $\cW$ is also given. The solution is then obtained by means of a collocation method with quadratic interpolation. Those works make use of the expression of the operator $|\partial_x|$ in the direct space. More recently, Karlin \textit{et al.} presented \cite{Movchan} a general iterative method for solving \eqref{PN}, based on the expression of $\dx\eta$ in the Fourier space. Our work borrows from the latter reference. The interested reader can also refer to \cite{BUEN14} for Fourier-based numerical schemes applied to the fractional Laplacian operator $\dx^\alpha$ with $\alpha>0$.

The present article proposes a numerical method to approximate solutions to \eqref{Wr} in the case where $F_\sigma$ is bistable. As in \cite{Movchan}, we build a dynamical system that admits \eqref{Wr} as its long-time limit, namely,
\begin{align}
\label{Wd2}
\lt\{
\begin{aligned}
    &\partial_t u(t,x)-c(t)\,\partial_x u(t,x)+ |\partial_x| u(t,x)= - F_\sigma'(u(t,x)),\\
    &u(t=0,x)=u_0(x),
\end{aligned}
\right.
\end{align}
where $x\in\mathbb{R}$, and $u_0$ is a regular initial data taking values in the interval $\lt[\eta_{\rm r},\eta_{\rm l}\rt]$ such that
\begin{align}
\label{DI}
    u_0(-\infty)&=\eta_{\rm l}\quad\text{and}\quad u_0(+\infty)=\eta_{\rm r},
\end{align}
where $\eta_{\rm l,r}$ are zeros of $F'_\sigma(\eta)$.

The iterative numerical approach introduced below uses (\ref{Wd2}) to approximate the solution of \eqref{Wr}. It is immediate that if $(\eta,\cW)$ solves \eqref{Wr} and if we impose $c(t):=\cW$, then $u(t,x):=\eta(x)$ satisfies \eqref{Wd2} for the initial data $u_0=\eta$. It is proved in~\cite{MathWeert} that if $F_\sigma$ is bistable, then \textit{for any} initial data $u_0$ with values in $\lt[\etar,\etal\rt]$ that satisfies \eqref{DI} and for \textit{any} continuous function $c(t)$, the solution $u$ of \eqref{Wd2} converges to the solution of \eqref{Wr} in a sense explained in Sec.\ \ref{SecConverge} below. This leads to the following procedure:
    \begin{enumerate}
      \item consider an initial condition $u_0(x)$ such that \eqref{DI} holds;
      \item approximate the solution $u(t,x)$ to \eqref{Wd2};
      \item while evaluating $u(t,x)$, choose $c(t)$ such that the core of $u(t,x)$ remains within the computational box and that $c(t)$ converges to $\cW$;
      \item for $t=t_{\text{f}}$ sufficiently large so that $u\lt(t,x\rt)$ has numerically converged, return $u\lt(t_{\text{f}},x\rt)$ as a numerical approximation to $\eta$, and deduce the velocity $\cW$.
    \end{enumerate}
As will be shown, this strategy proves efficient in cases of physical interest. However there might exist alternative strategies for solving \eqref{Wr}.

Numerical approximation of \eqref{Wr} paves the way to investigating numerically the Dynamic PN equation \cite{YPP}, which generalizes the Weertman equation to transient regimes. Indeed, the initial conditions and long-time steady-state regimes of the Dynamic PN equation are solutions to the latter equation \cite{YPP1}. However, we emphasize that~\eqref{Wd2} is only an algorithmic tool that has no relationship whatsoever with the actual dynamics of dislocations.

The article is organized as follows. In Sec.\ \ref{SecWeertman}, we briefly study the Weertman equation, and discuss both the uniqueness of its solution and its interpretation as the long-time limit of the dynamical system \eqref{Wd2}. We formally derive asymptotes of solutions to \eqref{Wr} and state identities about the velocity $\cW$ in general cases. Also, we explain how to choose $c(t)$ in \eqref{Wd2} to solve this equation in a \emph{comoving frame}, namely, one which follows the translational motion of the core. An analytical solution to \eqref{Wr} that exists in a particular case is recalled. In Sec.\ \ref{SecDisc}, we introduce our numerical representation for $\eta$ and discuss corresponding implementations of the diffusive operator $-|\partial_x|$ and the advective operator $\partial_x$, as well as methods to evaluate $c(t)$. Also, we make use of the asymptotic behavior when $|x| \rightarrow +\infty$ of the solution to \eqref{Wr} to circumvent the issue of the infinite domain of integration in \eqref{eq:hilb}. Once these fundamental elements have been introduced, we build in Sec.\ \ref{SecAlgo} a Preconditioned Collocation Scheme (PCS) that applies to our problem, and justify this denomination. In Sec.\ \ref{SecResNum}, we use this numerical approach on two test cases: one with a simple potential $F_\sigma$, for which the exact solution is known, and one with a more physically relevant potential $F_\sigma$. We also illustrate the robustness of our approach, concluding that the algorithm presented is unconditionally stable with respect to the time step $\Delta t$. We empirically derive error scalings with respect to the parameters involved in the discretization. A concluding discussion closes the article, underlining some limitations of our approach, and proposing a few possible extensions. An Appendix is devoted to examining further one such extension.

\section{Some properties of the Weertman equation}
\label{SecWeertman}
This section is devoted to an overview of some important properties of the Weertman equation \eqref{Wr} and of the companion dynamical equation \eqref{Wd2}.
\subsection{Invariances}
\label{sec:invar}
As the PN equation, Equation \eqref{Wr} is obviously invariant by translation. When invoking uniqueness of the solutions, we shall henceforth implicitly refer to `uniqueness up to arbitrary translations'. This invariance has consequences on the numerical solution, which usually undergoes an undesirable drift during calculations if no corrective action is undertaken. A special procedure is developed in Section \ref{SecCentering} below to eliminate this difficulty.

Moreover, Eq. \eqref{Wr} is invariant by reflection in the sense that if $(\eta(x),\cW)$ is a solution for boundary conditions $\eta(\pm\infty)=\eta_{\rm l,r}$, then  $(\eta(-x),-\cW)$ is another solution for boundary conditions $\eta(\pm\infty)=\eta_{\rm r,l}$. Therefore, without loss of generality, we always assume throughout this article that $\eta_{\rm l}>\eta_{\rm r}$.

\subsection{Existence and uniqueness of solutions to the Weertman equation.}
\label{sec:exist}
\begin{figure}[ht]
    \begin{center}
	\centering \includegraphics[width=8cm]{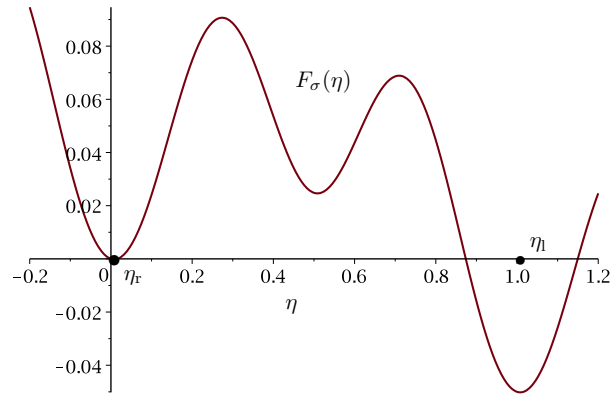}
    \end{center}
    \caption{\label{Fig.F} Camel-hump potential $F_{\sigma}$ defined by \eqref{Def_C-H}, with parameters $\sigma=0.05$ and $r=5$.}
\end{figure}

There exists a unique solution to \eqref{Wr} when $F_\sigma$ is a bistable nonlinearity. More precisely it can be shown rigorously (the proof relies on a recent result \cite{Gui}) that for $F_\sigma$ sufficiently regular, if $\eta_{\rm l}>\eta_{\rm r}$ are such that: (i) $F_\sigma'(\eta_{\rm l,r})=0$ and $F_\sigma''(\eta_{\rm l,r})>0$; (ii) any local minimizer $u$ of $F_{\sigma}$ between $\eta_{\rm r}$ and $\eta_{\rm l}$ satisfies $F_{\sigma}(u)>F_{\sigma}(\eta_{\rm r})$ and $F_{\sigma}(u)>F_{\sigma}(\eta_{\rm l})$, then there exists a unique velocity $\cW$ and a decreasing function $\eta$ satisfying \eqref{Wr}, which is unique up to translation. Condition (i) means that $F_\sigma$ is a `bistable potential'. A typical example of such $F_\sigma$, to be used in Sec.\ \ref{SecCamel}, is drawn in Fig \ref{Fig.F}. It corresponds to
\begin{align}
    \label{Def_C-H}
    &F_{\sigma}(\eta):= \frac{1}{4\pi} \lt[1-\theta^2-\lt(\theta \sqrt{1-\theta^2}+\arcsin(\theta)-\phi\rt)\cot\phi\rt]-\sigma\,\eta,\\
    &\theta=\sin(\phi)\cos(2\pi\eta),\qquad \phi=\arctan r.\nonumber
\end{align}
Possible non-decreasing solutions to \eqref{Wr} that might exist as in the PN equation \cite[Equation (16)]{Nabarro46}, e.g., for $\eta_{\rm l}=\eta_{\rm r}$, are disregarded in the present work. Hereafter, we always assume that $\eta_{\rm l,r}$ obey the above conditions, and we term such values of $\eta(x)$ at infinity \emph{consistent boundary conditions} (CBCs).

\subsection{Asymptotic behavior and characteristic lengths}
\label{SecAsymp}
By letting $|x|\to\infty$, we formally deduce the leading-order asymptotic expansions
\begin{align}
\label{Asympt1}
\eta(x)&\underset{x \rightarrow \pm\infty}\sim\eta_{\rm r,l}+\frac{\eta_{\rm l}-\eta_{\rm r}}{\pi F_{\sigma}''(\eta_{\rm r,l})} x^{-1}.
\end{align}
They are proved rigorously in \cite{MathWeert}. The key ingredient of the proof is the following asymptotic behavior of integrals with Cauchy kernel (see \cite[p.\ 267]{Musk}):
\begin{align}
\label{eq:asympt}
|\partial_x|\eta(x)&\sim\frac{1}{\pi x}\int_{-\infty}^{+\infty}\partial_x\eta(x')\dd x'=\frac{1}{\pi x}(\eta_{\rm r}-\eta_{\rm l}).
\end{align}
It can be formally retrieved from the leading term of a `series expansion' of \eqref{eq:hilb} at $x$ large (note however that the integral involved in the next-to-leading term of such a formal expansion may not exist). Substituting expression (\ref{eq:asympt}) into \eqref{Wr}, using the Taylor expansions at boundary values
\begin{align}
F'_\sigma(\eta)=F'_\sigma(\eta_{\rm l,r})+F''_\sigma(\eta_{\rm l,r})(\eta-\eta_{\rm l,r})+\ldots\simeq F''_\sigma(\eta_{\rm l,r})(\eta-\eta_{\rm l,r}),
\end{align}
and noting that $\partial_x\eta(x)$ vanishes like $O\lt(|x|^{-2}\rt)$ \cite{Gui}, we obtain \eqref{Asympt1}.

Introducing the characteristic lengths
\begin{align}
\label{eq:alrscales}
a_{\rm l,r}&=\frac{2\pi}{F_{\sigma}''(\eta_{\rm l,r})},
\end{align}
Equation \eqref{Asympt1} can be rewritten as
\begin{align}
\label{eq:asymptwitha}
\eta(x)&\underset{x \rightarrow \pm\infty}\sim\eta_{\rm r,l}+(\eta_{\rm l}-\eta_{\rm r})\frac{a_{\rm r,l}}{2\pi^2} x^{-1}.
\end{align}
It will be argued in Sec.\ \ref{SecSolAnalyt} that $a_{\rm l,r}$ represent typical scales of variation of the dislocation density on both sides of the solution. For further purposes, it is useful to introduce a mean characteristic length as
\begin{align}
\label{eq:ameanscale}
\overline{a}:=\frac{1}{2}(a_{\rm l}+a_{\rm r})=\pi\left[\frac{1}{F_{\sigma}''(\eta_{l})}+\frac{1}{F_{\sigma}''(\eta_{r})}\right].
\end{align}

Finally, depending on the potential $F_\sigma(\eta)$ at hand, the solution can be either \emph{symmetric}, in the sense that $\partial_x\eta(x)$ is an \emph{even} function (an example will be given in Sec.\ \ref{SecSolAnalyt}), or \emph{non-symmetric} if the latter property does not hold.

\subsection{Velocity determination}
\label{SecVitesse}
There are many ways to determine the velocity $\cW$ associated with the solution $\eta$ to Equation \eqref{Wr}. One possibility is to multiply \eqref{Wr} by a function $g$ chosen such that the integrals involved make sense, and to integrate over $\R$. This yields
\begin{align}
\cW&=\frac{\int_{-\infty}^{+\infty}g(x)\lt[|\partial_x|\eta(x)+F'_\sigma(\eta(x))\rt]\dd x}{\int_{-\infty}^{+\infty}g(x)\partial_x\eta(x)\,\dd x}.
\end{align}
In particular, the following choices of $g$ eliminate $|\partial_x|\eta(x)$. With $g(x)=1$ one gets
\begin{align}
\label{Idc2}
\cW&=\frac{\int_{-\infty}^{+\infty}F'_\sigma(\eta(x))\dd x}{\eta_{\rm r}-\eta_{\rm l}},
\end{align}
where the improper integral is understood as a principal value at infinity \cite[p.\ 252]{Gradshteyn}. Taking instead $g(x)=\partial_x\eta(x)$ \cite{Gui}, one arrives at an expression easier to implement
than \eqref{Idc2}, namely,
\begin{align}
\label{Idc1}
\cW&=\frac{F_\sigma(\eta_{\rm r})-F_\sigma(\eta_{\rm l})}{\int_{-\infty}^{+\infty}[\partial_x\eta(x)]^2 \dd x}.
\end{align}

\subsection{Convergence towards solutions to the Weertman equation}
\label{SecConverge}
The first author (M.J.) proves in \cite{MathWeert} that, under our working hypotheses, all the solutions to \eqref{Wd2} converge towards the unique solution of \eqref{Wr} at exponential rate. Indeed, consider the following equation:
\begin{align}
\label{Wd}
\partial_t \varphi(t,x)+ |\partial_x| \varphi(t,x)= - F_{\sigma}'(\varphi(t,x))\quad\text{with}\quad \varphi(t=0,x)=u_0(x),
\end{align}
where $u_0(x)$ is the initial condition of \eqref{Wd2}. The connection between \eqref{Wd} and \eqref{Wd2} resides in that $\varphi(t,x)$ solves \eqref{Wd} if and only if
\begin{align}
\label{eq:equivuv}
u(t,x)&=\varphi\left(t,x+\int_0^t c(s)\dd s\right)
\end{align}
solves \eqref{Wd2}. Now, under mild requirements similar to those of Sec.\ \ref{sec:exist}, Equation \eqref{Wd} can be shown \cite{MathWeert} to have a unique solution with the following property: if $(\eta,\cW)$ is the solution to \eqref{Wr} with same boundary conditions at infinity as $u_0$, then there exist constants $\kappa>0$, $K>0$ and $\xi \in \mathbb{R}$ such that
\begin{align}
\label{cv}
\sup_{x\in \mathbb{R}}|\varphi(t,x)-\eta(x-\cW t+\xi)|\leq K {\rm e}^{-\kappa t}.
\end{align}
The proof relies on a comparison principle, which is a generic property of operators $\partial_t - D + F_{\sigma}'\lt[\cdot\rt]$ where $D$ is a dissipative operator (see the classical references \cite{Fife, XinfuChen}).

Combining Equations \eqref{eq:equivuv} and \eqref{cv}, one deduces that at large times and uniformly in $x$,
\begin{align}
\label{eq:ulargetime}
u(t,x)&\simeq \eta(x+\zeta(t)),\quad\text{where}\quad \zeta(t)=\int_0^t c(s)\dd s-\cW t+\xi.
\end{align}
Thus, given CBCs at infinity, the long-time limit of the solution to \eqref{Wd2} is the solution to \eqref{Wr} with same CBCs, up to a time-dependent drift. In the next section, we show how a suitable choice of $c(t)$ eliminates this undesirable effect.

\subsection{Centering and choice of \texorpdfstring{$c(t)$}{c(t)}}
\label{SecCentering}
The present section essentially relies on formal arguments. To prescribe $c(t)$ in Equation \eqref{Wd2}, we need first to center at $x=0$ the core of the solution $\eta(x)$ of \eqref{Wr} by imposing the supplementary condition
\begin{align}
 \label{CondCenter}
  \frac{1}{2L}\int_{-L}^L \eta(x)\dd x =\overline{\eta},
\end{align}
where we have introduced the quantity
\begin{align}
\label{eq:etabar}
\overline{\eta}:=(\eta_{\rm l}+\eta_{\rm r})/2,
\end{align}
and where the constant $L>0$ represents the half-size of the computational box in the numerical calculations. Imposing condition \eqref{CondCenter} makes the solution of \eqref{Wr} unique (and not only unique up to translations), which is crucial for numerical purposes.

Next, on the basis of expression \eqref{Idc1} for $\cW$ , we prescribe the function $c(t)$ as
\begin{align}
\label{DefVitesseDyn2}
c(t):=\frac{F_\sigma(\eta_{\rm r})-F_\sigma(\eta_{\rm l})}{\int_{-\infty}^{+\infty}[\partial_xu(t,x)]^2\,\dd x}+\frac{\kappa}{(\etal-\etar)}I(t),\quad\text{where}\quad
I(t):=\int_{-L}^L \lt[u(t,x)-\overline{\eta}\rt]\dd x,
\end{align}
and $\kappa >0$ is a fixed parameter (the reciprocal of some characteristic time). Since  by \eqref{eq:ulargetime} $u(t,x)$ in \eqref{Wd2} converges to $\eta(x)$ up to a translation, we formally have by comparing the following expression to \eqref{Idc1}:
\begin{align}
\label{cdyn11}
\frac{F_\sigma(\eta_{\rm r})-F_\sigma(\eta_{\rm l})}{\int_{-\infty}^{+\infty}[\partial_xu(t,x)]^2\,\dd x}\underset{t \rightarrow+\infty}{\longrightarrow} \cW.
\end{align}
Substituting this limit into definition (\ref{DefVitesseDyn2}) one deduces that at large times
\begin{align}
\label{eq:ctlargetime}
c(t)\simeq \cW+\frac{\kappa}{(\etal-\etar)} I(t).
\end{align}

If we can show that $I(t)\to 0$ when $t\to\infty$, which is equivalent to
\begin{align}
\label{eq:centeringu}
\frac{1}{2L}\int_{-L}^L u(t,x)\,\dd x&\to\overline{\eta}
\end{align}
by definition of $I(t)$, then in the same limit we shall have by \eqref{eq:ctlargetime} $c(t)\to\cW$. The limit \eqref{eq:centeringu} indicates that the choice \eqref{DefVitesseDyn2} forces the dynamical solution $u(t,x)$ to obey asymptotically the same centering as $\eta(x)$. Put differently, this amounts to computing $u(t,x)$ in a comoving frame. The rightmost term in \eqref{DefVitesseDyn2}${}_1$ can thus be called a \emph{centering correction} to the velocity.

To show that $I(t)\to 0$, we differentiate the large-time expression \eqref{eq:ulargetime}${}_1$ of $u(t,x)$ with respect to time. Further invoking \eqref{eq:ctlargetime}, we obtain at large times (the dot denotes a total time derivative)
\begin{align}
\label{eq:partialtu}
\partial_t u(t,x)
&\simeq \partial_x\eta(x+\zeta(t))\dot{\zeta}(t)=\partial_x\eta(x+\zeta(t))[c(t)-\cW]
\simeq \partial_x\eta(x+\zeta(t))\frac{\kappa}{(\etal-\etar)}I(t).
\end{align}
Integrating \eqref{eq:partialtu} over $x\in[-L,L]$ then yields the approximate first-order differential equation
\begin{align}
\label{eq:Dyn_Sur_I}
\dot{I}(t)
&\simeq -\kappa \frac{\eta(-L+\zeta(t))-\eta(L+\zeta(t))}{\etal-\etar}I(t)\simeq -\kappa\,I(t),
\end{align}
where the rightmost expression follows from having neglected the $1/x$ term in the asymptotic expansions \eqref{eq:asymptwitha} of $\eta(x)$. From \eqref{eq:asymptwitha} this is legitimate if $L\gg \max(a_{\rm l},a_{\rm r})/(2\pi^2)+|\zeta(t)|$. The latter condition is compatible with $L$ being fixed in numerical computations if $\zeta(t)$ tends to a finite value at large times. The latter property follows from a simple self-consistent argument. Indeed, Equation \eqref{eq:Dyn_Sur_I} implies that $I(t)\simeq I(T)\ee^{-\kappa(t-T)}$ for $t>T$ where $T$ is some time above which \eqref{eq:ulargetime} holds, so that effectively $I(t)\to 0$. Substituting this expression of $I(t)$ into \eqref{eq:ctlargetime}, one deduces an approximate analytical expression for $c(t)$ that tends to $c_\eta$. Writing \eqref{eq:ulargetime}${}_2$ in the form $\zeta(t)=\zeta(T)+c_\eta(T-t)+\smash{\int_T^t} c(s)\,\dd s$, and substituting the obtained expression of $c(t)$ finally entails the desired saturation property in the form $\zeta(t\to\infty)\simeq \zeta(T)+I(T)/(\eta_{\rm l}-\eta_{\rm r})$. Unfortunately no quantitative estimate of the latter quantity is available. Yet, on the basis of numerical experiments, it can be expected to be of the order of the (unknown) dislocation core width.

As a final remark, we anticipate by indicating that, later on, $\kappa$ will be taken inversely proportional to the algorithmic time step. Thus, in practice $\kappa\to\infty$ in the limit of continuous times. Because of the $\exp(-\kappa t)$ dependence of $I(t)$, we observe that the centering correction in \eqref{DefVitesseDyn2}${}_1$ remains well-behaved in this limit.

\subsection{An analytical solution}
\label{SecSolAnalyt}
For $|\sigma|<1$ and
\begin{equation}
\label{Fsin}
F_{\sigma}'(\eta)=\sin\lt(2\pi \eta\rt)-\sigma,
\end{equation}
with CBCs
\begin{align}
\eta_{\rm r}&=\arcsin(\sigma)/(2\pi),\quad\text{and}\quad \eta_{\rm l}=1+\eta_{\rm r},
\end{align}
the dimensionless Equation \eqref{Wr} admits the following analytical solution, easily deduced from the well-known solution \cite{Rosakis} to the original Weertman equation:
\begin{subequations}
\begin{align}
\label{EquAna1}
&\eta(x)=\eta_{\rm r}+\frac{\eta_{\rm l}-\eta_{\rm r}}{\pi}\lt[\frac{\pi}{2}-\arctan\lt(\frac{2\pi x}{a} \rt)\rt]
\quad\text{and}\quad
\cW(\sigma)=\tan\lt(2\pi\eta_{\rm r}\rt)=\frac{\sigma}{\sqrt{1-\sigma^2}},\\
\label{EquAna2}
\text{with}&\quad a=1/\cos\lt(2\pi\eta_{\rm r}\rt)=\frac{1}{\sqrt{1-\sigma^2}}.
\end{align}
\end{subequations}
This solution is symmetric in the sense of Section \ref{SecAsymp}. We will use this test case in Sec.\ \ref{SecResNum} as a benchmark. So, when $|\sigma|\to 1$ the core width $a$ and the velocity $\cW$ both blow up as
\begin{align}
\label{eq:blowup}
a(\sigma)&\simeq1/\sqrt{2\lt(1-|\sigma|\rt)}\quad\text{and}\quad \cW(\sigma)\simeq{\rm{sgn}}(\sigma)/\sqrt{2(1-|\sigma|)}.
\end{align}
Since $\cW$ and $a$ are not the physical velocity and core width, this behavior is not the hallmark of a physical pathology of the model. It however implies that the comptational box should be taken wider and wider to contain the core of $\eta$, and that the latter moves with nearly infinite velocity, which has numerical consequences to be examined in Sec.\ \ref{sec:convprops}.

In the form \eqref{eq:asymptwitha}, the asymptotic behaviors deduced from a direct expansion of (\ref{EquAna1})${}_1$ read
\begin{align}
\label{eq:exasympt}
\eta(x)&\mathop{\sim}_{x\to\pm\infty}\eta_{\rm r,l}+(\eta_{\rm l}-\eta_{\rm r})\frac{a}{2\pi^2 x},
\end{align}
where the length scales \eqref{eq:alrscales} are $a_{\rm l}=a_{\rm r}=a$. Thus, in this particular example where the solution is symmetric the asymptotic behaviors provide a connection between the core width and the next-to-leading terms in the expansion. This supports the interpretation put forward in Sec.\ \ref{SecAsymp} that in generic asymmetric situations $a_{\rm l,r}$ represent characteristic scales of variation of the dislocation density.

\section{Building blocks}
\label{SecDisc}
Our numerical scheme to solve the dynamical system \eqref{Wd2} crucially rests on evaluating the action of the operator $|\partial_x|$ in the Fourier domain, by means of the Discrete Fourier Transform (DFT). This section explains the underlying spatial discretization procedure, and outlines the key features of the implementation.
\subsection{Temporal and spatial discretization}
\label{sec:tempspadis}
At discrete times $t_n=n\Delta t$ with step $\Delta t>0$, we need a suitable representation $U^n(x)$ of $u(t_n,x)$ over the whole $x$-axis. To this aim, we define a computational box $[-L,L]$, discretized into $2m$ elementary intervals of width $h=L/m$. We write the function $U^n(x)$ as
\begin{align}
\label{Decompo}
 U^n(x)&=V^n(x)+\etaref(x),
\end{align}
where $V^n(x)$ is a time-evolving part, the support of which is contained \emph{within} the box $[-L,L]$, and $\etaref(x)$ is a fixed reference function that complies with the asymptotic behaviors \eqref{Asympt1}. The latter function is prescribed once for all, and plays the role of boundary conditions for the operators $|\partial_x|$ and $\partial_x$. Decomposition \eqref{Decompo} is motivated by the fact that the operator $\dx$ is non-local, and that the solution $\eta$ of \eqref{Wr} does not vanish at infinity. Thus, the tail contributions of $u(t,x)$ at infinity should be taken into account when computing $\dx u$. They are represented in \eqref{Decompo} by those of $\etaref(x)$. The need to properly account for tail contributions will be illustrated by means of numerical examples in Section \ref{SecParamAlgo}.

In this article, we take $\etaref$ as the linear combination
\begin{subequations}
\label{Formulas_Etaref}
\begin{align}
\label{eq:etaref}
\etaref(x)&=\sum_{\alpha=1}^4 A_\alpha f_\alpha(x/\aref),
\end{align}
where the basis functions $f_\alpha(x)$ are chosen such that $\dx f_\alpha(x)$ can be computed analytically, as
\begin{align}
f_1(x)&:=1,&&
\text{with}\quad
\dx f_1(x)=0,\\
f_2(x)&:=-\frac{1}{\pi}\arctan\lt(2\pi x\rt),&&
\text{with}\quad
\dx f_2(x)=-\frac{4\pi x}{4\pi^2 x^2+1},\\
f_3(x)&:=\frac{x}{x^2+1},&&
\text{with}\quad
\dx f_3(x)=\frac{2x}{(x^2+1)^2},\\
f_4(x)&:=\frac{1}{\sqrt{x^2+1}},&&
\text{with}\quad
\dx f_4(x)=\frac{2}{\pi}\frac{x\ln\lt[(x^2+1)^{1/2}-x\rt]+\sqrt{x^2+1}}{\lt(x^2+1\rt)^{3/2}},
\end{align}
\end{subequations}
and where $\aref$ is an extra arbitrary scaling parameter. The four coefficients $A_\alpha$ are determined so as to satisfy the four constraints expressed by \eqref{Asympt1}. Comparing the asymptotic expansions of $\etaref(x)$ for $x\to\pm\infty$ directly deduced from \eqref{Formulas_Etaref} to the generic expressions \eqref{Asympt1} leads to
\begin{align}
&A_1=\overline{\eta},\quad
A_2=\eta_{\rm l}-\eta_{\rm r},\quad
A_3=\frac{A_2}{2\pi^2}\left(\frac{\overline{a}}{\aref}-1\right),\quad
A_4=\frac{A_2}{2\pi^2}\frac{\overline{a}}{\aref}\frac{F''_\sigma(\eta_{\rm l})-F''_\sigma(\eta_{\rm r})}{F''_\sigma(\eta_{\rm l})+F''_\sigma(\eta_{\rm r})},
\end{align}
where $\overline{a}$ and $\overline{\eta}$ and have been defined, respectively, in Equations \eqref{eq:ameanscale} and \eqref{eq:etabar}. Thus, choosing $\aref=\overline{a}$ makes $A_3$ vanish. In the exact case of Sec.\ \ref{SecSolAnalyt}, where furthermore $F''_\sigma(\eta_{\rm l})=F''_\sigma(\eta_{\rm r})$, only $A_1$ and $A_2$ are nonzero and the solution is already completely retrieved at the level of $\etaref(x)$. Therefore we shall need to take $\aref$ different from $\overline{a}$ to be able to use the exact solution of Sec.\ \ref{SecSolAnalyt} as a non-trivial benchmark of the algorithm in Sec.\ \ref{SecResNum}. We observe that the asymptotic expansion of $\etaref(x)$ in (\ref{Formulas_Etaref}) involves only \emph{odd} inverse powers of $|x|$.

Finally, introducing discrete positions $x_j=j\,h$, the function $U^n(x)$ is represented \emph{inside} the box by the vector $\mathbf{u}^n \in \mathbb{R}^{2m}$ of components $u^n_j=U^n(x_j)$. It is decomposed according to \eqref{Decompo} as
\begin{align}
\label{eq:unj}
u^n_j = v^n_j+ \etaref(x_j), \quad \text{for} \quad  j \in \{-m,\cdots,m-1\},
\end{align}
where the vector $\mathbf{v}^n$ of components $v^n_j$ is now the unknown of the problem.

It must be emphasized that, however convenient, representation \eqref{Formulas_Etaref} is largely arbitrary. Indeed, any other smooth function $\etaref(x)$ obeying the asymptotic conditions \eqref{Asympt1}, mostly varying inside the computational box, and such that $\dx \etaref(x)$ can be accurately computed once for all (either analytically, or even numerically), would equally well fit our purpose.

\subsection{Numerical Fourier transform and discretization of the diffusion operator}
\label{SecNumFour}
\label{SecDiff}
This section addresses the discretization of the linear integro-differential operator $|\partial_x|$. In view of \eqref{eq:hilb}, the latter involves a convolution by a pseudofunction \cite{KANW04}, which can be done in the Fourier representation. Crucially, the present approach uses the \emph{continuous} Fourier form of the operator because this representation is straightforwardly diagonal, and versatile in the sense that it could as well be employed to address the fractional Laplacian $|\partial_x|^\alpha$.

The FT will be implemented in DFT form, which allows one to benefit from Fast-Fourier-Transform (FFT) routines. As is well known, carrying out a convolution over the real axis by means of DFT and multiplications in the Fourier domain turns this convolution into a periodic one, which induces undesirable periodicity effects in the direct space near both extremities of the interval of interest \cite{Bracewell}.

The usual workaround is to use zero-padding. Briefly, the technique consists in doubling the spatial extent of the  interval of interest, in such a way that the undesirable effects that take place during convolution remain confined to the extra region. The vector to be convolved is continued by zero in the latter region. Once the convolution with the kernel has been carried out in the Fourier space by Fourier component multiplication, and the result has subsequently been transformed back to direct space by DFT inversion, the irrelevant contributions of the added region that have been spoiled by periodicity effects are dropped. This last step constitutes a projection of the result onto the initial interval of interest.

For details the reader is referred to the classical reference \cite[p.\ 643]{NumRecipes}. However, the latter reference only addresses situations where the initial vector and the convolution kernel are both known in the direct space. In contrast, since the present approach uses the convolution kernel in continuous Fourier form, it requires identifying the Fourier modes $k$ in the continuum to the discrete modes $k_p$ used by the DFT. For that reason, the labelling and ordering of the vector components hereafter somewhat differ from the ones in Ref.\ \cite{NumRecipes}. However, as Equation \eqref{eq:conv} below shows, the final outcome will take the form of a standard convolution in the direct space.

Thus, zero-padding is carried out by means of the injection $\mathcal{I}$ that maps $\mathbb{R}^{2m}$ into $\mathbb{R}^{4m}$, and the inverse transform involves a projector $\mathcal{P}$ that maps $\mathbb{R}^{4m}$ onto $\mathbb{R}^{2m}$. These operators are defined by
\begin{align}
\label{eq:defcali}
&\mathcal{I} \lt(u_0,\cdots,u_{m-1},u_{-m},\cdots,u_{-1} \rt)
=\lt(u_0,\cdots,u_{m-1},0,\cdots,0,u_{-m},\cdots,u_{-1}\rt),\\
\label{eq:defcalp}
&\mathcal{P}\lt(v_{0},\cdots,v_{m-1},\cdots,
v_{2m-1},v_{-2m},\cdots,v_{-m},\cdots,v_{-1} \rt)
=\lt(v_0,\cdots,v_{m-1},v_{-m},\cdots,v_{-1} \rt).
\end{align}
Note that $\mathcal{P} \mathcal{I}$ leaves $\mathbb{R}^{2m}$ invariant.

The DFT that operates on extended vectors $\mathbf{v}\in \mathbb{R}^{4m}$ is denoted hereafter by $\Fd\lt\{\mathbf{v}\rt\}$ (or, alternatively, by $\Fd\lt\{v_p\rt\}$). The corresponding enlarged spatial grid that extends the one in  Equation \eqref{eq:unj} is $x_j$ for $j\in \mathcal{K}:=\{-2m,\cdots,2m-1\}$. It admits the dual wavemode grid
\begin{align}
& k_p:=2\pi p/(4 L), \quad\text{for}\quad p\in\mathcal{K}.
\end{align}
Accordingly, DFTs are carried out for each wave mode $p$ as
\begin{align}
\left(\Fd\lt\{\mathbf{v}\rt\}\right)_p:=\sum_{j=-2m}^{2m-1} v_j {\rm e}^{-{\rm i} x_j k_p} = \sum_{j=-2m}^{2m-1}v_j {\rm e}^{-{\rm i}  \frac{2\pi  j  p}{4m}}.
\end{align}

From \eqref{Decompo} we can now compute the discretized form of $\dx U(x_j)$ as
\begin{align}
\label{eq:udisc}
\dx U(x_j) &=\dx V(x_j)+\dx \etaref(x_j),
\end{align}
in which $\dx \etaref$ is evaluated at point $x_j$ by means of \eqref{Formulas_Etaref}, and $\dx V$ is computed via the DFT as
\begin{align}
\label{eq:vdisc}
\dx V(x_j)\simeq\left(\projec\Fd^{-1}\lt\{|k_p|\lt(\Fd \lt\{\injec \mathbf{v}\rt\}\rt)_p\rt\}\right)_j.
\end{align}
Combining \eqref{eq:udisc} and \eqref{eq:vdisc} we shall introduce $\dxnum \mathbf{u}$, the discretization of $\dx U$, defined as
\begin{align}
\label{Defdxnum}
\lt(\dxnum \mathbf{u}\rt)_j = \dx \etaref(x_j)+\lt(\projec \Fd^{-1} \lt\{ |k_p|  \lt(\Fd \lt\{  \injec \mathbf{v}\rt\}\rt)_p  \rt\}\rt)_j,\qquad j \in \lt\{-m,\cdots,m-1\rt\}.
\end{align}
A straightforward check shows that the rightmost term simplifies as
\begin{align}
\lt(\Fd^{-1} \lt\{ |k_p| \lt(\Fd \lt\{ \mathcal{I} \mathbf{v}\rt\}\rt)_p  \rt\}\rt)_j
\label{eq:conv}
&=\sum_{l=-m}^{m-1}\lt(\Fd^{-1}\lt\{\lt|k_p\right|\rt\} \rt)_{j-l} v_{l}.
\end{align}
Although there are $4m$ wave modes, the sum only involves indices $l\in\{-m,\cdots,m-1\}$, precisely because of the use of injection $\mathcal{I}$. As desired, the convolution in (\ref{eq:conv}) is well-behaved and not spoiled by undesirable periodic effects, since it distinguishes at each point $x_j$ the contributions $l<j$ of the $u_l$ on the left of $x_j$, from those on the right for $l>j$. This would not have been the case, had fewer points been retained to compute DFTs.

\subsection{Discretization of the advection operator}
\label{SecAdv}
 We discretize the advection operator $c\partial_x$ with the following upwind scheme \cite{Fletcher}:
\begin{align}
\label{eq:AdvDef}
 c\partial_x U(x_j) \simeq \lt(\lt[c^-D_-+ c^+ D_+\rt] \lt\{U(x_l) \rt\}_l\rt)_j, \qquad j \in \{-m,\cdots,m-1\},
\end{align}
where $c^+:=\max\lt(c,0\rt)$ and $c^{-}:=\min\lt(c,0\rt)$, and where the operators $D_{\pm}$ are implemented via expressions with $\bigO(h^3)$ error (see \cite[p.\ 297]{Fletcher} with $q=1/2$), namely,
\begin{subequations}
\label{eq:thirdorderadv}
\begin{align}
\lt( D_- \mathbf{u}\rt)_j&:= \frac{2 u_{j+1} +3 u_j - 6 u_{j-1}+u_{j-2}}{6 h},\\
\lt( D_+ \mathbf{u}\rt)_j&:= \frac{-u_{j+2}+6 u_{j+1} - 3 u_{j}-2 u_{j-1}}{6 h}.
\end{align}
\end{subequations}
The choice a third-order implementation is motivated in Sec.\ \ref{SecParamAlgo} by numerical considerations (see Ref.\ \cite{Fletcher} for schemes of orders $1$ and $2$ and Ref.\ \cite[p.\ 111]{Hildebrand} for order $4$). Since
\begin{align}
U(x_l)=\lt(\injec\mathbf{v}\rt)_l+\etaref(x_l), \qquad l \in \KK,
\end{align}
the discretization \eqref{eq:AdvDef} of the advection operator actually reads
\begin{align}
\label{eq:DiscAdv}
c\pxnum[c] \mathbf{u}:=\projec \lt(\lt[c^-D_-+ c^+ D_+\rt] \lt\{\lt(\injec\mathbf{v}\rt)_l+\etaref(x_l)  \rt\}_l\rt),
\end{align}
where $\pxnum[c]$ only depends on the sign of $c$. In this expression, $\etaref$ plays the role of boundary conditions when computing $\partial_x U$ at the extremities of the computational box, e.g., for $l\in \{-m,-m+1,m-2,m-1\}$ in \eqref{eq:DiscAdv}.

Note that whereas, in principle, implementing $D_{\pm}$ should require adding only a few extra points at the exterior of the interval $[-L,L]$, our implementation \eqref{eq:DiscAdv} is carried out in practice on $4m$ points, due to the introduction of $\mathcal{I}$ and $\mathcal{P}$ for consistency with the implementation of the operator $\dx$. Also, remark that the operators $D_{\pm}$ are almost diagonal, in the sense that their periodizations are diagonal in the Fourier space. In this connection, we introduce for further use in Section \ref{SecPCS} the vectors $\mathbf{D}_{\pm} \in \mathbb{R}^{4m}$ defined as
\begin{align}
\label{eq:DefD}
\lt(\mathbf{D}_{\pm}\rt)_{j}:= \lt(D_{\pm} \mathbf{e}_0\rt)_{-j},
\end{align}
with $\mathbf{e}_0=(1,0,\cdots,0) \in \mathbb{R}^{4m}$, and where the periodic convention $0=4m$  applies in the evaluation of the operator $D_{\pm}$ in \eqref{eq:DefD}.

\subsection{Velocity computation}
\label{SecVitesseNum}
As seen above, Equation \eqref{Wd2} can be solved in the comoving frame by using the velocity $c(t)$ given by \eqref{DefVitesseDyn2}. In this way, the dislocation core lies as remote as possible from the box boundaries to minimize the influence of the approximations made in handling the tails. To proceed, we substitute in \eqref{eq:DiscAdv}, at each time $t_n$, the quantity $c$ by $c_n=c(t_n)$ (with the convention that $c_{-1}=0$) computed from a discretized version over $2m-1$ points of expression \eqref{DefVitesseDyn2}, in which $L$ has been replaced by $L-h$ (since $x_{-(m-1)}=-L+h$ and $x_{m-1}=L-h$); namely,
\begin{subequations}
\begin{align}
\label{DefVitesseDyn3}
c_n
&:=[F_\sigma(\eta_{\rm r})-F_\sigma(\eta_{\rm l})]\lt[h\sum_{j=-(m-1)}^{m-1}\omega_j \lt|\lt(D_x\lt[c_{n-1}\rt]\bm{u}^n\rt)_j\right|^2+\frac{\eta_{{\rm l},1}^2+\eta_{{\rm r},1}^2}{3(L-h)^3}\rt]^{-1}
+\frac{\kappa}{\eta_{\rm l}-\eta_{\rm r}}I_n,\\
I_n&:= h\sum_{j=-(m-1)}^{m-1} w_j\lt(u^n_j - \overline{\eta}\rt).
\end{align}
\end{subequations}
In these expressions $\eta_{{\rm l\,r},1}=(\eta_l-\eta_r)a_{\rm l,r}/(2\pi^2)$, and $\omega_j$ are the weights $1/3$, $4/3$, $2/3$, \ldots, $2/3$, $4/3$, $1/3$ of the Simpson integration rule. Remark that $c_n$ depends only on $c_{n-1}$ via its sign, which is of little consequence except in calculations at $\sigma$ small where $c_n$ is close to $0$ and may oscillate during iterations. The term within brackets results from a straightforward discretization of the integral in \eqref{DefVitesseDyn2}${}_1$, in which the tail contributions have been evaluated analytically from the asymptotic expansions of $\etaref$ (for simplicity, we have not evaluated the full tail contributions of $\etaref$).

Moreover, a suitable value of $\kappa$ stems from the empirical consideration that if we discretize Equation \eqref{eq:Dyn_Sur_I} in explicit Euler form as $I_{n+1}=(1-\kappa \Delta t)I_n$, monotone convergence towards 0 of $I_n$ is ensured by taking $\kappa < 1/\Delta t$. Correspondingly, the value of $\kappa$ used henceforth in \eqref{DefVitesseDyn3} is $\kappa=1/(2\Delta t)$.

\section{Algorithm}
\label{SecAlgo}
This section describes the iterative numerical scheme used to compute $\bm{\eta}$ and $c_{\bm{\eta}}$. This algorithm is explicit in time, is consistent with the dynamical system \eqref{Wd2}, combines the above-discretized operators, and remains stable even with a large time step $\Delta t$.
\subsection{Procedure}
Computations go as follows. First, for given local minimizers $\eta_{\rm l}$ and $\eta_{\rm r}$ of $F_{\sigma}$, the algorithm is typically initialized by choosing arbitrarily the values $u^0_j$ inside the box, preferentially not too far from the expected solution. This can be done in a number of ways; notably, by using as an initial condition the function $\smash{\etaref}$ with $\smash{\aref}$ chosen large enough to encompass the expected overall width of the dislocation density (typically, a few times the characteristic scale $\overline{a}$), whence $u^0_j\equiv 0$. A few low-resolution runs may help adjusting $\smash{\aref}$. Obviously, to get a reasonable representation of the solution, the discretization step $h$ must necessarily be less than $\overline{a}$. Also, when carrying out incremental parametric studies, the solution computed from the previous value of the parameter under consideration can be used as an initial condition, to save CPU time. However, for studying stability issues, we shall purposely take initial data far from the expected shape of a dislocation.

Denoting by $\Phi$ the scheme presented below, we iterate
\begin{align}
\mathbf{u}^{n+1} = \Phi\lt(\mathbf{u}^n\rt)
\end{align}
until the difference between the results of two successive iterations is small, in the sense that
\begin{equation}
\label{ConvIndidc}
\lt\| \Delta \mathbf{u}^n\right\|:=\max_{j\in\{-m,\cdots,m-1\}} \lt|u_j^n-u_j^{n+1}\right|\leq\Delta_0  \Delta t,
\end{equation}
where $\Delta_0$ is a user-defined stopping criterion. Upon completion at some $n$, the algorithm returns $\bm{\eta}:=\mathbf{u}_n$ and the associated $c_{\bm{\eta}}=c_n$, evaluated thanks to \eqref{DefVitesseDyn3}. Unless otherwise stated, $\Delta_0=10^{-10}$ in the numerical calculations of Sec.\ \ref{SecResNum} below.

\subsection{The Preconditioned Collocation Scheme}
\label{SecPCS}
The scheme $\Phi$ described hereafter, which we call the \emph{Preconditioned Collocation Scheme} (PCS), is based on the requirement that the long-time limit $\bm{\eta}$ of $\mathbf{u}^n$ should solve the following static equation:
\begin{align}
\label{BasePrecond}
-\lt(\dxnum \bm{\eta}\rt)_j +c\lt(\pxnum[c] \bm{\eta}\rt)_j=F_\sigma'(\eta_j), \qquad j \in \{-m,\cdots,m-1\},
\end{align}
which is the discretized form of \eqref{Wr}. This is a collocation method. A first naive way to proceed would be to attempt solving \eqref{Wd2} by writing
\begin{align}
\label{Precond}
\mathbf{u}^{n+1}=\mathbf{u}^n + \Delta t \big[ -\dxnum \mathbf{u}^n + c_n\pxnum[ c_n] \mathbf{u}^n-F_\sigma'\lt(\mathbf{u}^n\rt)\big],
\end{align}
where $\Delta t$ should be adjusted in order to achieve convergence. At convergence, the solution $\bm{\eta}$ obeys \eqref{BasePrecond} and depends on $\Delta t$ only through the evaluation of the numerical velocity $c_n$ defined by \eqref{DefVitesseDyn3}. As will be justified in Section \ref{SecParamDisc} this dependence is of little relevance, which is an advantage of this approach.

However, system \eqref{Precond} is ill-conditioned, because it involves the stiff operators $-\dxnum$ and $\pxnum[c_n]$. As a consequence, \eqref{Precond} does not converge if $\Delta t$ is not small enough. This issue is dealt with by preconditioning \eqref{Precond} in the following way:
\begin{subequations}
\label{eq:precondall}
\begin{align}
\label{Precond2}
\mathbf{u}^{n+1}&=\Phi(\mathbf{u}^n)=\mathbf{u}^{n} + \Delta t\,M(\Delta t)\lt\{ -\dxnum \mathbf{u}^n +  c_n\pxnum\lt[c_{n}\rt] \mathbf{u}^n-F_\sigma'\lt(\mathbf{u}^n\rt)\rt\},
\end{align}
where the $\mathbf{u}^n$-dependent operator $M(\Delta t)$ acts on a vector $\mathbf{u}\in\mathbb R^{2m}$ as
\begin{align}
\label{eq:Precond}
M(\Delta t)\mathbf{u}&=\mathcal{P} \Fd^{-1}\lt\{ \frac{\lt(\Fd\lt\{\mathcal{I} \mathbf{u}\rt\}\rt)_p}{1+\Delta t \lt|k_p\right| -\Delta t \lt(\Fd\lt\{c_n^+ \mathbf{D}_++c_n^- \mathbf{D}^- \rt\}\rt)_p }\rt\}.
\end{align}
\end{subequations}
With the PCS \eqref{eq:precondall} the long-time limit of $\mathbf{u}^n$ also satisfies Equation \eqref{BasePrecond}. Therefore, the preconditioning achieved by introducing $M(\Delta t)$ is just a procedure to enable and speed up convergence.

Moreover, if we ignore the operators $\mathcal{I}$ and $\mathcal{P}$ in \eqref{eq:Precond}, the matrix $M(\Delta t)$ is close to the effective inverse of $1+\Delta t \dxnum -\Delta t c_n \pxnum[c_n]$. Hence, \eqref{Precond2} preconditioned by \eqref{eq:Precond} is nothing but the following semi-implicit scheme (see \cite[p.\ 102]{HairerWanner2} for a reference on semi-implicit schemes):
\begin{align}
\label{eq:SIS}
\mathbf{u}^{n+1}\simeq\mathbf{u}^n +
\Delta t\lt[-\dxnum\mathbf{u}^{n+1} + c_n\pxnum\lt[c_n\rt] \mathbf{u}^{n+1} - F_\sigma'\lt(\mathbf{u}^n\rt)\rt].
\end{align}
As is well-known, treating stiff operators in an implicit way yields a stable scheme. Hence, this preconditioning naturally leads to stability (this assertion will be exemplified in Section \ref{SecResNum}). We note that \eqref{eq:SIS} is a consistent discretization of \eqref{Wd2} ---just as \eqref{Precond}.

We now justify the preconditioned character of $\Phi$ by analogy with the task of solving iteratively a linear system. Indeed, ignoring nonlinearities by replacing $F_\sigma'\lt(\mathbf{u}^n\rt)$ by a constant $\mathbf{b}$ and by setting $c_n=0$, \eqref{BasePrecond} can be put in the form $\mathsf{A}\mathbf{u}=\mathbf{b}$, so that \eqref{Precond} reduces to a scheme of the type
\begin{align}
\mathbf{u}^{n+1} = \mathbf{u}^n - \Delta t\lt(\mathsf{A}\mathbf{u}^n -\mathbf{b}\rt),
\end{align}
where $\mathsf{A}$ is a \emph{positive} symmetric matrix. The latter scheme converges for general $\mathbf{b}$ if and only if all eigenvalues of $\lt(1-\Delta t\,\mathsf{A}\rt)$ belong to the interval $(-1,1)$; this can require $\Delta t$ to be very small. In a similar way, the scheme \eqref{eq:precondall} can be abstracted as
\begin{align}
\label{eq:abstrast2}
\mathbf{u}^{n+1}=\mathbf{u}^n-\Delta t \mathsf{M}\lt(\mathsf{A}\mathbf{u}^n -\mathbf{b}\rt),
\end{align}
with $\mathsf{M}$ close to $\lt(1+\Delta t\mathsf{A}\rt)^{-1}$. Then, the eigenvalues of the latter scheme are close to those of $(1+\Delta t\,\mathsf{A})^{-1}$, which unconditionally belong to $(0,1)$ if $\Delta t>0$. Equation \eqref{eq:abstrast2} amounts to solving $\mathsf{M}\mathsf{A}\mathbf{u}=\mathsf{M}\mathbf{b}$, which is the classical preconditioning method (in the spirit of a modified Richardson iteration, see \cite[p.\ 6]{Kelley}).

\section{Numerical results}
\label{SecResNum}
The above algorithm has been implemented in the form of a {\sc Matlab}\textsuperscript{\textregistered} code, and the following results have been obtained on a 2.3GHz standard laptop computer, with typical computation times of order one second to a few minutes per run, depending on the case at hand. Except in Sec.\ \ref{SecCamel}, the calculations concern benchmark comparisons with the exact solution of Sec.\ \ref{SecSolAnalyt}, and have been carried out with $F'_{\sigma}$ as defined by \eqref{Fsin}, for $\sigma \in [0,1)$. Moreover, in Sections \ref{sec:convprops} to \ref{SecParamAlgo}, a parameter value $\aref=\overline{a}/2$ (see Section \ref{sec:tempspadis}) is employed in $\smash{\etaref(x)}$.

\subsection{Convergence}
\label{sec:convprops}
\begin{figure}[!h]
\centering \includegraphics[width=17cm]{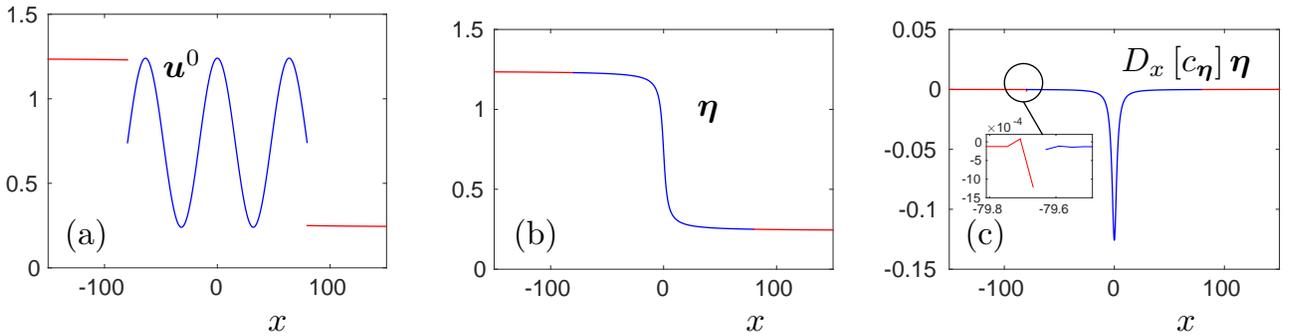}
\caption{\label{FigConverge}(a) Initial data $\mathbf{u}^0$; (b) output $\bm{\eta}$; (c) discrete derivative of $\bm{\eta}$. Blue: solution $\bm{\eta}$; red: parts of $\etaref$ outside the box. Discretization parameters $2m=4096$, step $h=L/m\simeq 0.39$, and $\Delta t=0.1$.}
\end{figure}
As stated in Sec.\ \ref{SecSolAnalyt}, the core width $a(\sigma)$ and the velocity $\cW(\sigma)$ blow up when $\sigma \rightarrow 1$. The first problem is easily solved by running one preliminary low-resolution run to provide a rough numerical estimate for the core width $\widetilde{a}(\sigma)$. The half computational box size $L$ is then adjusted to a value $L\gg \widetilde{a}(\sigma)$. Figure \ref{FigConverge} displays the initial data $\mathbf{u}^0$, the converged result $\bm{\eta}$ and its discrete derivative. The figure illustrates the robustness of the PCS with respect to initial conditions in the sense that the initial data $\mathbf{u}^0$ can be non-monotone, irregular, and far from the solution $\eta$ to \eqref{Wr}. In this calculation, the applied loading is $\sigma=1-1.973\times 10^{-3}$, which induces large values for the converged velocity and core width, $\cW(\sigma)\simeq a(\sigma)\simeq 15.9$, close to one another; see Equation \eqref{eq:blowup}. The half box size is $L=5\,a(\sigma)\simeq 80$. As seen in Fig.\ \ref{FigConverge}(c) the discretized derivative of $\bm{\eta}$ is regular inside the box $[-L,L]$, but the inset shows that some artifacts take place near the matching points between the solution inside the box and the tails of $\etaref(x)$ outside it. Quite generally, it is observed that these artifacts diminish when either $L$ or $\sigma$ are increased (results not shown).

\begin{figure}[ht]
\begin{center}		
\centering \includegraphics[height=5.5cm]{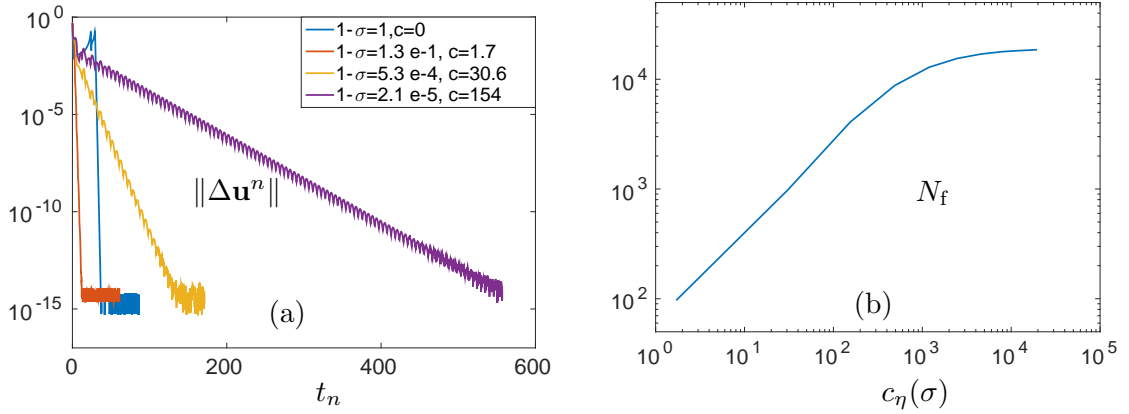}
\end{center}
\caption{\label{Fig.Cv} (a) Convergence indicator $\lt\|\Delta \mathbf{u}^n\right\|$ defined by \eqref{ConvIndidc} vs.\ time $t_n$, and (b) number of iterations $N_{\rm{f}}$ until $\lt\|\Delta \mathbf{u}^n\right\| \leq \Delta_0$ vs.\ velocity $\cW(\sigma)$. Parameters $L=10\,a(\sigma)$, $2m=1024$, and $\Delta t=0.1$.}
\end{figure}
Under the same initial conditions, Figure \ref{Fig.Cv} illustrates the convergence properties with time, via $\lt\|\Delta \mathbf{u}^n\right\|$ defined in \eqref{ConvIndidc}. The semi-logarithmic plot of Fig.\ \ref{Fig.Cv}(a) shows that, up to high-frequency oscillations, convergence is exponential with time, in agreement with the arguments of Section \ref{SecConverge}. Moreover, the convergence rate (the slopes in Fig.\ \ref{Fig.Cv}(a)) depends on the applied loading $\sigma$, convergence being impeded when $\sigma$ approaches $1$. Not unexpectedly, this loss of performance coincides with $F_{\sigma}$ being `less and less bistable' in the sense that the minima of $F_\sigma$ at $\eta=\eta_{\rm r}$ and $\eta=\eta_{\rm l}$ become less and less deep. Bistability is a crucial requirement for the existence of a solution to \eqref{Wr}, and for proving convergence as expressed by \eqref{cv}. In this connection, it should be remarked that the tail expressions \eqref{Asympt1} involve the second derivatives $F''_\sigma(\eta)$, which presumably leads to pathologies when the latter are small. However, Fig. \ref{Fig.Cv}(b), which displays the data in parametric form of parameter $\sigma$, shows that the PCS copes well with high velocities, which is important from a physical standpoint. Thus, our preconditioning does a nice job of avoiding stability issues when $\cW$ is high.

\subsection{Error indicators and overall accuracy}
\label{sec:errindic}
\begin{figure}[ht]
\centering
\includegraphics[height=5.5cm]{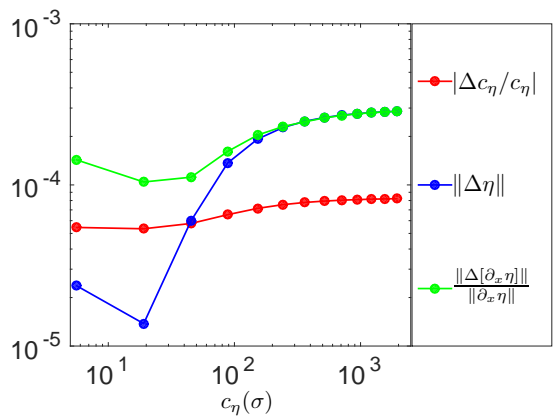}
\caption{\label{Fig_HautC}
Parametric plot with $\sigma$ as the variable parameter, of the error indicators \eqref{eq:errorindicators} vs.\ $\cW(\sigma)$. Parameters $L=20a(\sigma)$, $2m=2048$ and $\Delta t=0.1$.}
\end{figure}

Fig.\ \ref{Fig_HautC} compares the output of the PCS with the exact solution of Sec.\ \ref{SecSolAnalyt} in terms of the following indicators:
\begin{subequations}
\label{eq:errorindicators}
\begin{align}
\label{eq:Erroreta}
&\lt\|\Delta \eta\right\|:=\inf_{\xi \in \mathbb{R}} \max_{j\in\{-m,\cdots,m-1\}} \lt| \bm{\eta}_j - \eta(x_j+\xi) \right|,\\
\label{eq:Errordx}
&\frac{\lt\|\Delta \lt[\partial_x \eta\rt]\right\|}{\lt\|\partial_x \eta \rt\|}
:=\inf_{\xi \in \mathbb{R}} \max_{j\in\{-m,\cdots,m-1\}} \frac{\lt| \pxnum[\cnum]\bm{\eta}_j - \partial_x \eta(x_j+\xi) \right|}{\lt\|\partial_x \eta\rt\|_{{\rm L}^\infty(\R)}},\\
\label{eq:Errorc}
&\lt|\Delta \cW/\cW\right|:=|c_{\bm{\eta}}/\cW-1|.
\end{align}
\end{subequations}
The absolute error \eqref{eq:Erroreta} can as well be understood as a relative error since in the cases considered $\lt\|\eta\rt\|_{{\rm L}^{\infty}(\R)}$ is of order $1$. In \eqref{eq:Erroreta} and \eqref{eq:Errordx} the $\inf_{\xi}$ operation, motivated by the translation invariance of \eqref{Wr}, takes care of the approximate character of the centering of the numerical solution, due to discretization errors. Formulated in this way, the error indicators are insensitive to small shifts in the position of the computed solution. The figure indicates that the outputs of the PCS accurately approximate the exact results, with errors of a similar order of magnitude for the three quantities represented. In addition, the small errors observed on Figure \ref{Fig_HautC} depend only weakly on $\cW$.

\subsection{Discretization parameters and error scaling}
\label{SecParamDisc}
This section closely investigates the scalings of the error at convergence with respect to the time step $\Delta t$, the half box size $L$ and the space discretization step $h$. The same case as above is considered, with a variety of applied loadings.

\begin{table}[ht]
\caption{\label{Plusieurdt}Errors as a function of $\Delta t$ ($2L=638$, $2m=4096$, $h=0.156$, $\sigma_1=0.9921$).}
\centering
\begin{tabular}{|c|c|c|c|c|c|c|c|c|}
\hline
$\Delta t$              & 0.25                  & 0.1                     & 0.01                & 0.001\\
\hline
$\lt\|\Delta \eta\rt\|$ & $5.39992\times 10^{-5}$ & $5.39991\times 10^{-5}$  & $5.39990\times 10^{-5}$ & $5.39990\times  10^{-5}$\\
\hline
$\lt\|\Delta c/c\rt\|$ & $2.01277\times 10^{-5}$ & $2.01266\times  10^{-5}$     & $2.01260\times  10^{-5}$ & $2.01260\times  10^{-5}$\\
\hline
\end{tabular}
\end{table}
Addressing first the influence of $\Delta t$, we observe that the PCS solution depends on $\Delta t$ only through the centering correction term in expression \eqref{DefVitesseDyn3} of $c_n$. However, as explained in Section \ref{SecVitesseNum}, this term vanishes in the limit of infinite times, and is therefore expected to be small at convergence. This is confirmed by the errors reported in Table \ref{Plusieurdt} for applied loading $\sigma=\sigma_1$ (see caption) so that $\cW(\sigma_1)\simeq 7.91$, and decreasing values of $\Delta t$. Errors are quasi-constant, which shows that the dependence on $\Delta t$ of the converged result is negligible. For definiteness, the rest of the calculations in the present paragraph is made with $\Delta t=0.1$.

\begin{figure}[ht]
\centering
\includegraphics[height=5cm]{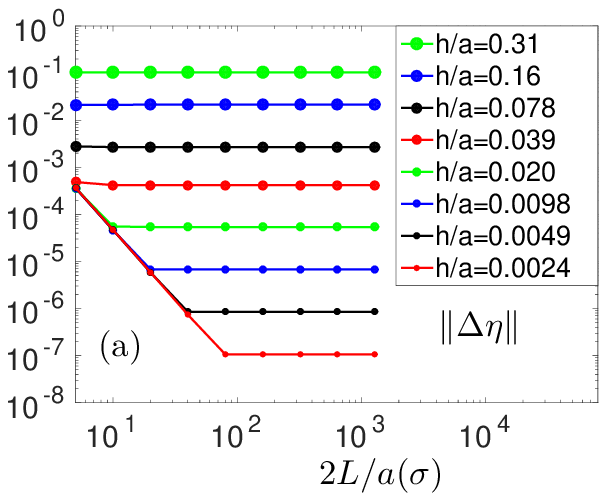}\includegraphics[height=5cm]{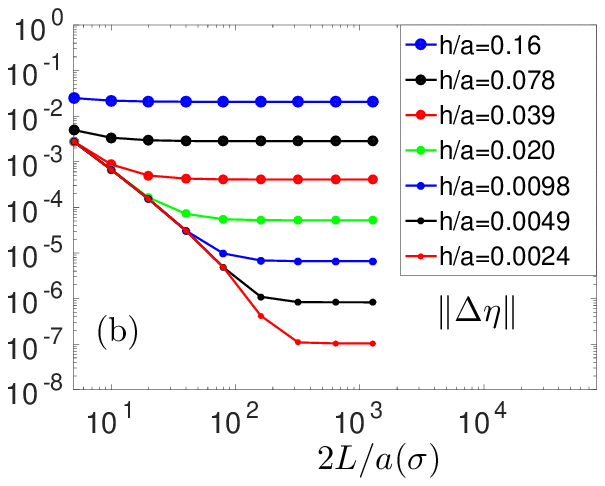}\\
\includegraphics[height=5cm]{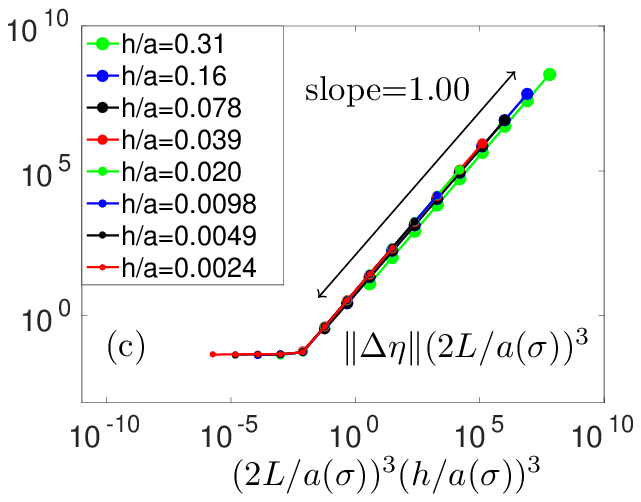}\includegraphics[height=5cm]{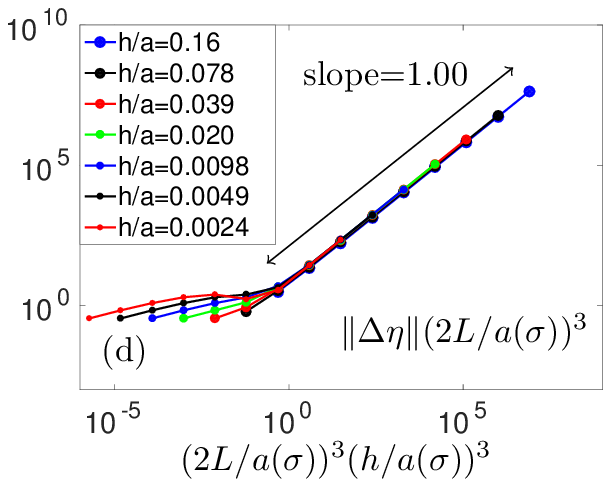}
\caption{\label{Multihdt}
Raw data and corresponding data-collapse plots for error $\left\|\Delta \eta\right\|$. Loading $\sigma=\sigma_2$ in (a) and (c) (moderate velocity), and $\sigma=\sigma_3$ in (b) and (d) (high velocity); see text. In the legends, $a$ stands for $a(\sigma)$.}
\end{figure}
Figure \ref{Multihdt} illustrates how $\left\|\Delta \eta\right\|$ depends on $L$ and $h=L/m$, for the two contrasted velocities obtained with loadings $\sigma_2=0.951$ so that $\cW(\sigma_2)\simeq 3.08$, and $\sigma_3=1-1.97\times 10^{-5}$ so that $\cW(\sigma_3)\simeq 159$. Raw results are displayed in Figs.\ \ref{Multihdt}(a) and (b). Computations for $\sigma=\sigma_3$ did not converge with $h/a=0.31$, which is why this value is not considered in plots (b) and (d). It turns out that for $L$ large, $\left\|\Delta \eta\right\|$ scales as $h^{3}$ for both loadings. Besides, it is approximately proportional to $L^{-3}$ for $L$ small. This is demonstrated in the data-collapse plots (c) and (d) where the individual datasets of Fig.\ \ref{Multihdt}(a) and (b), respectively, are merged into one single master curve by means of appropriate rescalings of abscissas and ordinates. The data collapse of Fig.\ \ref{Multihdt}(b) for $\sigma=\sigma_3$ is only partially successful, as noticeable corrections to scaling arise for $Lh\ll a(\sigma_3)^2$.

The scalings can be understood as follows. On the one hand, the scaling in $h$ is consistent with our choice of a third-order advection scheme in \eqref{eq:thirdorderadv}. On the other hand, since the error at any point is spread over the whole domain by the integro-differential operator $\dx$, the error $\eta(L)-\etaref(L)$ at the boundary point $x=L$ (for instance) can be used to estimate the overall error. It behaves as $L^{-3}$ in the present case where $\eta(x)$ and $\etaref(x)$ are symmetric, and where the next nonzero term in expansion \eqref{eq:asymptwitha} is $\propto x^{-3}$. This is consistent with the error scaling observed in the plots. Still, these elementary arguments do not explain the downwards bending of the high-velocity plots in Fig.\ \ref{Multihdt}(b), which indicates either that the $L^{-3}$ scaling regime has not been reached, or that it may not hold exactly. This bending causes deviations from ideal scaling, made conspicuous in Fig.\ \ref{Multihdt}(d).

\begin{figure}[ht]
\centering
\includegraphics[height=5cm]{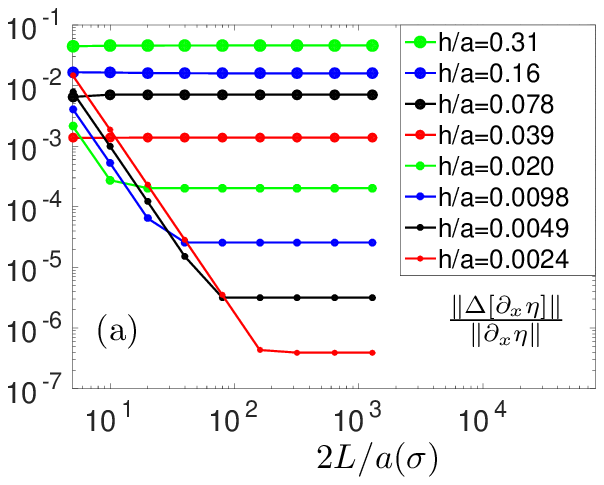}
\includegraphics[height=5cm]{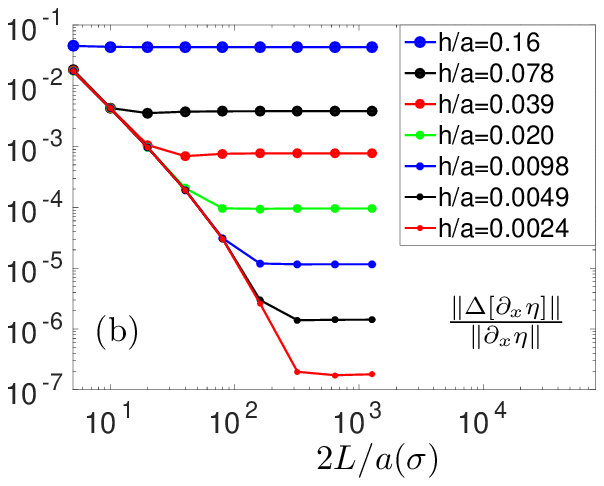}
\caption{\label{Qual_hautC}
Raw data for the error \eqref{eq:Errordx} on $\partial_x\eta(x)$. (a) $\sigma=\sigma_2$, (b) $\sigma=\sigma_3$ (see text).}
\end{figure}
\begin{figure}[ht]
\centering
\includegraphics[height=5cm]{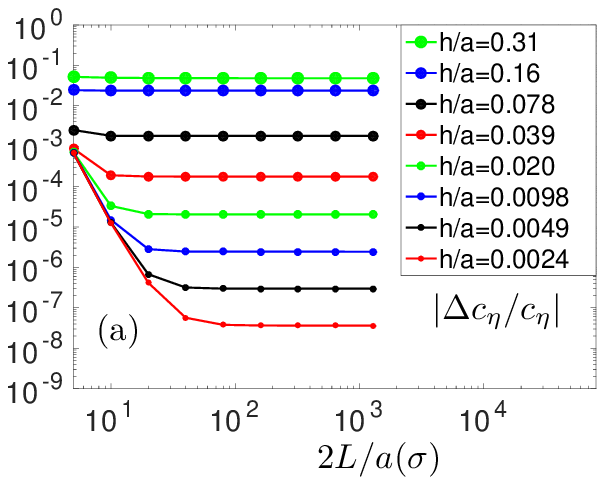}
\includegraphics[height=5cm]{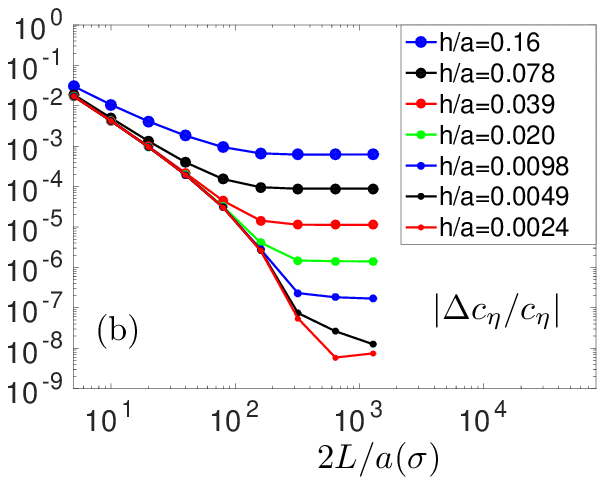}\\
\caption{\label{ERC_Der}
Raw data for the error \eqref{eq:Errorc} on $\cW$. (a) $\sigma=\sigma_2$, (b) $\sigma=\sigma_3$ (see text).}
\end{figure}
Likewise, Figures \ref{Qual_hautC} and \ref{ERC_Der} display the errors \eqref{eq:Errordx} and \eqref{eq:Errorc}, respectively, for the loadings $\sigma_2$ and $\sigma_3$. Whereas Fig.\ \ref{Qual_hautC}(a) resembles Fig.\ \ref{Multihdt}(a), the dependence of the error with respect to $h$ is more involved. In particular for $\sigma=\sigma_2$ (Fig.\ \ref{Qual_hautC}(a)), there exists at fixed $L$ an optimal value of $h=L/m$ that minimizes the error. The reason for this behavior is unclear. However it should be realized that approximating  $\partial_x \eta$ in the sense of the ${\rm L}^{\infty}$ norm is quite demanding. Figs.\ \ref{Qual_hautC}(b) and \ref{ERC_Der}(b) display bendings similar as in Fig.\ \ref{Multihdt}(b). No data collapse is presented, as notable deviations from scaling take place in all figures.

\begin{table}[ht]
\caption{\label{TableCost} Error on $\eta$ minimized over $L$ for $\sigma=\sigma_2$, as a function of $m$.}
\centering
\begin{tabular}{|c|c|c|c|c|c|c|c|}
\hline
$2m$&$32$&$128$&$512$&$2048$&$8192$\\
\hline
min.\ value of $\lt\|\Delta \eta\rt\|$&$2.1 \times 10^{-2}$&$4.9\times 10^{-4}$&$ 5.5\times 10^{-5}$&$6.8 \times 10^{-6}$&$8.5 \times 10^{-7}$\\
\hline
optimum $hL/a(\sigma_2)^2$&$0.781$&$0.195$&$0.195$&$0.195$&$0.195$\\
\hline
\end{tabular}
\end{table}
For practical matters one needs, e.g., to determine the optimal value of $L$ at fixed number $2m$ of discretization points, which somehow corresponds to a fixed cost to go from $\bm{u}^n$ to $\bm{u}^{n+1}$ in \eqref{Precond2}. Table \ref{TableCost} shows the minimal error $\lt\|\Delta \eta\rt\|$ deduced from the datasets of Fig.\ \ref{Multihdt}, together with the corresponding optimal value of $L$ via the ratio $(hL)/a^2=(L/a)^2/m$. One observes that the optimal value of the latter quantity does not depend on $m$ for $m$ sufficiently large. This can be understood from the above scaling arguments. Indeed, the data collapse in Fig.\ \ref{Multihdt}(c) indicates that for $L h\ll C$, where $C$ is some constant, $\lt\|\Delta \eta\rt\|\propto L^{-3}$, and that for $L h\gg C$, $\lt\|\Delta \eta\rt\|\propto h^3=L^3/m^3$. Thus the optimum takes place at the crossover between these regimes in which the error first decreases, then increases with $L$. Balancing the two regimes, it follows that $L h=L^2/m\simeq C$ at the optimum. The analysis still qualitatively holds for higher $\sigma$ values. The optimum just discussed is not the one evoked above in connection with Fig.\ \ref{Qual_hautC}. However, in both cases, one sees that increasing the number of discretization points at fixed $L$ does not necessarily improve the accuracy: one also needs to increase $L$.

Same arguments as above suggests that, in the general case of a nonsymmetric solution where the next nonzero term in expansion \eqref{eq:asymptwitha} can be an $\bigO(x^{-2})$ (see Sec.\ \ref{SecAsymp}), the error on $\eta$ would behave as $L^{-2}$ instead of $L^{-3}$, and if the error scales as $h^3$ for $h$ small the optimum would take place for $L h^{3/2}\simeq C$.

\subsection{Influence of the tails and of the order of the advection scheme}
\label{SecParamAlgo}
At given $L$ the quality of the numerical solution also depends on how tail contributions are accounted for while handling the operator $\dx$. As the quantity $\dx\eta(x)$ enters the equation for $\eta$, errors on the former affect the latter via the nonlinear term. To illustrate this point, we focus on the point $x=0$ where $\eta(0)=\etaref(0)=\overline{\eta}$. Since both the function $\etaref$ and the solution $\eta$ to \eqref{Wr} satisfy the same asymptotic behavior \eqref{Asympt1}, their difference behaves as $\eta(x)-\etaref(x)=O\lt(|x|^{-\tau}\rt)$, where $\tau=2$ in the general case, and $\tau=3$ in the present, symmetric, benchmark case. Hence by \eqref{eq:hilb2}, \emph{the long-range contribution to the error} in $\dx\eta(0)$ due to the replacement of $\eta$ by $\etaref$ outside the computational box is bounded by
\begin{align}
\label{eq:errL3}
\lt|\int_{L}^{+\infty}\frac{\eta(y)+\eta(-y)-[\etaref(y)+\etaref(-y)]}{y^2}\dd y\rt|\lesssim\int_{L}^{+\infty}\frac{\dd y}{y^{\tau+2}}\propto L^{-(\tau+1)}.
\end{align}
Thus, in the results of the previous section \ref{SecParamDisc} for which $\tau=3$ the observed scaling stems from local errors at boundary points, which scale as $L^{-3}$, and not from the present long-range contributions of errors in tails, which scale at most as $L^{-4}$. In the general non-symmetric case where $\tau=2$, the scalings of these errors are presumably changed into $L^{-2}$ (see end remark in the previous section) and $L^{-3}$, respectively. Therefore, using a function $\etaref(x)$ with faithful tails (in the sense of Section \ref{sec:tempspadis}) should make local errors dominant over tail errors in all cases. In contrast, upon not using such a reference function, or upon periodizing the calculation (thus, disregarding tails in both cases), the error on $\dx\eta(0)$ would instead scale as
\begin{align}
\label{eq:errL4}
\left|\int_L^{+\infty} \frac{\eta(y)+\eta(-y)}{y^2}\dd y\right|\lesssim\int_{L}^{+\infty}\frac{\dd y}{y^2}=L^{-1}.
\end{align}
We indeed obtained such a scaling while carrying out some preliminary studies (results not shown). To summarize, disregarding tails deteriorates the accuracy of the calculations. On the contrary, appealing to the zero-padding trick and using a reference function $\etaref$ with faithful tails to handle properly the operator $\dx$ strongly improves it.

The overall error also depends on the order in $h$ of the discrete advection operator $D_{\pm}$ of Sec.\ \ref{SecAdv}. Along with the calculations of the previous Section with an advection scheme of order 3, we also carried out similar calculations with schemes of order 2 and 4. In both latter cases, the power of the scaling in $h$ was found identical to the order of the scheme for $h$ small enough (results not shown). Thus, at least for orders 2 to 4, the scaling in $h$ is closely related to the order or the discretization scheme of the advection operator, the discretization errors involved by the DFT in the computation of the operator $\dx$ presumably being subdominant.

\begin{figure}[ht]
\centering
\includegraphics{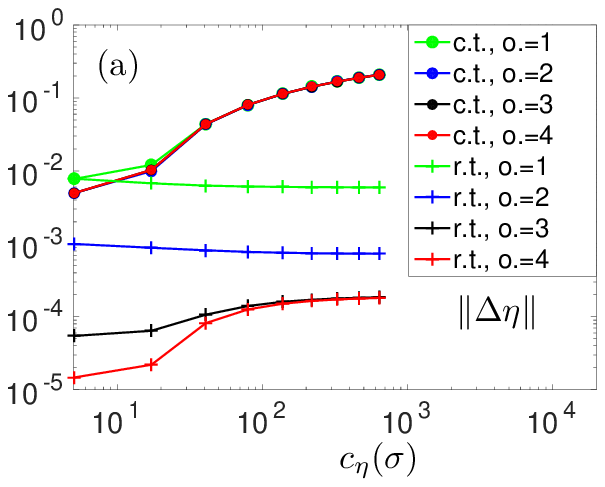}
\includegraphics{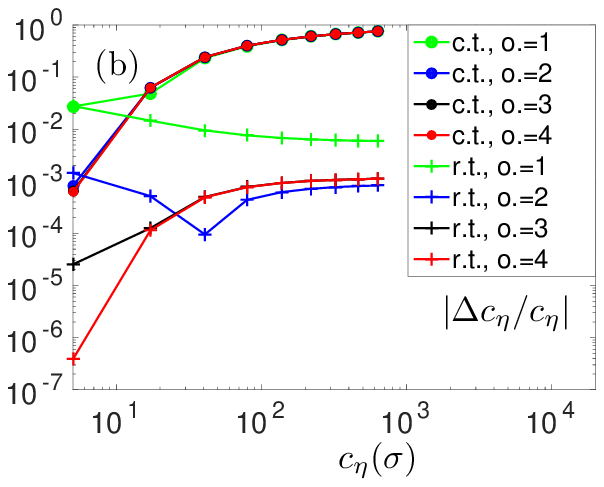}
\caption{\label{Contribords}
Parametric plots with $\sigma$ as the variable parameter. (a) $\lt\|\Delta \eta\right\|$, (b) $\lt|\Delta \cW/\cW\right|$  vs.\ $\cW(\sigma)$. Legends: $\rm{o.}$: order of the advection scheme; $\rm{r.t.}$  `refined tails'; ${\rm c.t.}$ 'constant tails' (see text). Discretization parameters $L=10 a(\sigma)$, $2m=1024$. The plots for $\rm{o.}=2$, $3$, and $4$ cannot be distinguished for constant tails.}
\end{figure}
Figures \ref{Contribords} further illustrate these points by means of the indicators \eqref{eq:Erroreta} and \eqref{eq:Errorc}. The plots have been made with advection upwind schemes of order $1$, $2$, $3$ and $4$, and two different types of asymptotes; namely, `refined tails' (r.t.) \eqref{Asympt1}, and `constant tails' (c.t.) in which the inverse power-law correction in \eqref{Asympt1} is dropped. Both figures show that using refined tails provides in most cases orders-of-magnitude gains of accuracy over using constant tails. The dependence of the error on the order of the upwind scheme is more difficult to interpret, although orders 3 and 4 lead to better accuracy. As the second-order scheme behaves irregularly in Fig.\ \ref{Contribords} (b), the third-order scheme is used in the rest of the article.

\subsection{A generalized example: the camel-hump potential}
\label{SecCamel}
We now evaluate our method with the `camel-hump' type potential $F_\sigma$ defined by \eqref{Def_C-H}. Depending on $\sigma$, and on the parameter $r>0$, $F_\sigma$ can feature between its two main minima at $\eta_{\rm l}$ and $\eta_{\rm r}$, and its humps, an intermediate local minimum of depth controlled by $r$, leading thus to a more or less dissociated solution $\eta$ to \eqref{Wr}. The potential \eqref{Def_C-H} has been derived by means of Lej\v{c}ek's method \cite{Lejcek75} so as to provide the following \emph{exact} dissociated dislocation solution to the PN equation when $\sigma=0$:
\begin{align}
\label{eq:solchsigma0}
\eta(x)&=\frac{1}{2\pi}\lt[\pi-\arctan\lt(2\pi x-r\rt)-\arctan\lt(2\pi x+r\rt)\rt]
\end{align}
(note that for $r=0$, the camel-hump potential reduces to the simple sinusoidal form \eqref{Fsin}).

\begin{figure}[ht]
\centering \includegraphics{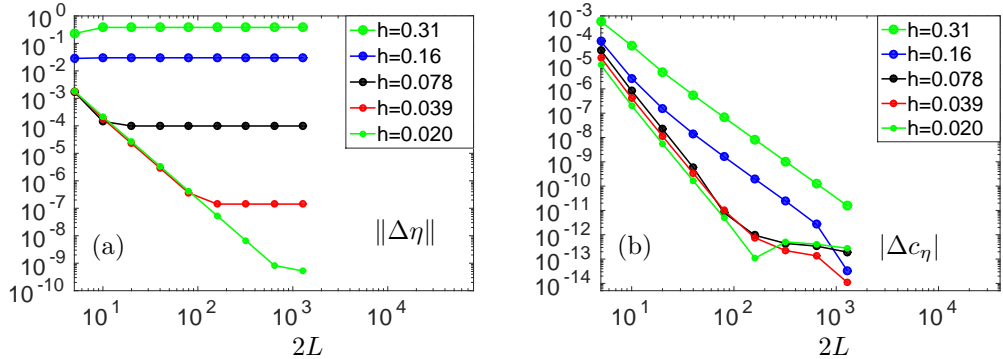}
\caption{\label{Fig_CMh}
(a) $\lt\|\Delta \eta\rt\|$ vs.\ $2L$; (b) $\lt|\Delta \cW\rt|$ vs.\ $2L$, for $F_\sigma$ as in \eqref{Def_C-H}. Parameters $r=5$, and  $\sigma=0$.}
\end{figure}
In this section, calculations have been carried out with $\aref=5\overline{a}$, because the overall width of the dislocation is much larger than that of the individual peaks in the density. Figure \ref{Fig_CMh} shows that the algorithm correctly recovers this solution. Due to the presence of a secondary hollow in the potential, the method takes longer to converge.

\begin{figure}[ht]
\centering \includegraphics{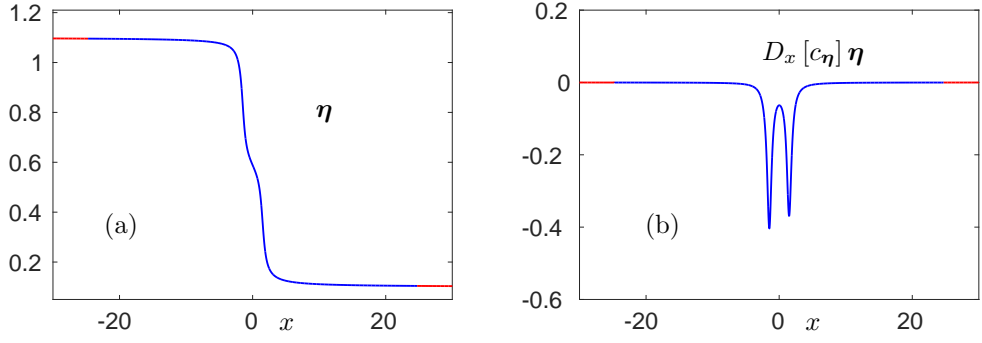}
\caption{\label{Fig.Eta_CH} (a) Numerical solution $\bm{\eta}$ and (b) discrete derivative of $\bm{\eta}$ for $F_\sigma$ defined by \eqref{Def_C-H} with parameters $r=5$ and $\sigma=0.5274$. Blue: solution $\bm{\eta}$; red: parts of $\etaref$ outside the box. Discretization parameters: $2L=49$ and $2m=4096$.}
\end{figure}
\begin{figure}[ht]
\centering
\includegraphics{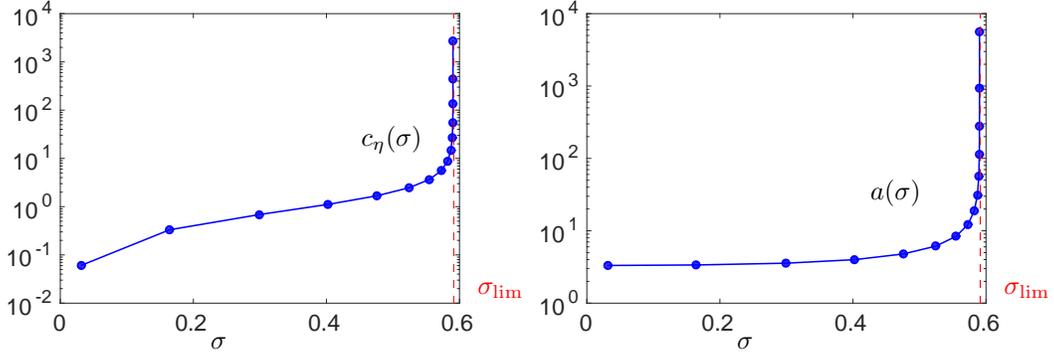}
\caption{(a) Velocity $\cW(\sigma)$, (b) core width $a(\sigma)$ for $F_\sigma$ defined by \eqref{Def_C-H} with $r=5$. Discretization parameters: $2m=2048$ and $L\simeq 20 a(\sigma)$.\label{DerFigure}}
\end{figure}
We finally present an application to a more physically relevant case, for which the exact solution is unknown. Indeed, when $\sigma>0$ and $r>0$  no analytical solution to \eqref{Wr} is available. However, as discussed in Sec.\ \ref{sec:exist}, there exists a solution to \eqref{Wr} that can be computed with our algorithm for any $\sigma\in(0,\sigma_{\text{lim}})$, where $\sigma_{\text{lim}}=\max_{\eta} F'_\sigma(\eta)$. Figure \ref{Fig.Eta_CH} displays the solution $\bm{\eta}$ (a) and its derivative (b), which is the dislocation density. In this example, the latter features two bumps that represent partial dislocations. The midpoint between the two partial dislocations corresponds to the local minimum of $F_{\sigma}$ in $\eta_{\rm m} \in \lt(\eta_{\rm r}, \eta_{\rm l}\rt)$. The asymmetry of the dislocation density can be interpreted as a consequence of the nonzero driving force $\sigma$ coupled to the dissociation process induced by the camel-hump character of the potential. Figure \ref{DerFigure} displays two quantities of physical interest, namely, the velocity $\cW(\sigma)$ and the effective core width $a(\sigma)$, for $\sigma$ varying between $0$ and $\sigma_{\text{lim}}=\max \lt(F_0'\rt)=0.5902$. Here, the quantity $a(\sigma)$ has been computed by minimizing the ${\rm{L}}^2$ norm of the difference between the numerical solution $\bm{\eta}$ and the function, parametrized by $a$ and $x_0$:
\begin{align}
f_{a,x_0}(x):=\eta_{\rm r}+\frac{\eta_{\rm l}-\eta_{\rm r}}{\pi} \lt[ \frac{\pi}{2} - \arctan\lt( \frac{2\pi (x-x_0)}{a} \rt) \rt].
\end{align}
As expected from an analogy with the simpler case of Section \ref{SecSolAnalyt}, both $a(\sigma)$ and $\cW(\sigma)$ increase with $\sigma$ and blow up when $\sigma\rightarrow\sigma_{\text{lim}}$.

\section{Concluding discussion}
To summarize, we have proposed the Preconditioned Collocation Scheme (PCS), which is a numerical procedure to approximate solutions to \eqref{Wr}, based on the dynamical system \eqref{Wd2}. The PCS uses the continuous FT of the operator $|\partial_x|$ and takes advantage of the FFT in its implementation, \emph{in spite of the strong constraint that the desired solution is not periodic, but has boundary conditions at infinity.} We have taken advantage from the exact asymptotic expansion of the solution to improve the accuracy of the numerical approximation. Also, we have shown that an overall $\bigO(h^{3})$ error in the space discretization could be achieved by means of a third-order advection scheme.

The method employed remains stable when the (a priori unknown) advection part scaled by the velocity $c_\eta$ dominates over the diffusion part in \eqref{Wr}, which allows to investigate the asymptotic behavior of \eqref{Wr} when $c_\eta$ is large. The PCS solves a discretized version of \eqref{Wr}. Being preconditioned, it can be used with a large time step $\Delta t$, nonetheless delivering outputs that depend very weakly on $\Delta t$. Although this was not illustrated, we add that if $\eta_{\rm l}$ and $\eta_{\rm r}$ are not exactly computed as exact local minimizers of $F_{\sigma}$ (e.g., if they suffer from slight numerical errors), the algorithm converges as well. Our method has some limitations, however: the more advection dominates diffusion, the more iterations the method requires to converge.

Still, the time and space complexities of the algorithm give satisfactory accuracy at reasonable computational cost. Indeed, the PCS requires $\bigO(m)$ memory space. Moreover DFTs have been speeded up by means of Fast Fourier Transform routines. Therefore, each iteration step takes $\bigO(m\log m)$ CPU time. Given that the number of iterations to convergence obviously scales as $1/\Delta t$, the PCS therefore has an overall time complexity of order $\bigO(m\log m/\Delta t)$. Then, achieving $10^{-4}$ accuracy on $\eta$ and $\cW$ with 2048 discretization points takes about one second on a standard laptop with CPU running at 2.3GHz, except in difficult cases when $\cW \rightarrow +\infty$, where it is slower. This study involved runs with up to $10^5$ discretization points.

We point out that alternative schemes could be used to simulate the dynamical system \eqref{Wd2}. First, other time integrators can be appealed to. For instance, classical splitting schemes such as Strang or Lie splittings \cite{Hairer2010} aim at simulating a dynamical system that involves a sum of operators, and turn the latter into a composition of evolution operators. Although these investigations were not reported for conciseness, we have checked that such methods do indeed apply to our problem, and prove stable and robust. However, the solution they produce is inherently $\Delta t$-dependent. Hence, achieving high accuracy requires small $\Delta t$ values, which makes them expensive. Also, as mentioned in Section \ref{SecPCS}, an explicit Euler scheme could as well be employed. However, the latter approach proves unstable if $\Delta t/h$ is not small. In contrast, the alternative semi-implicit scheme sketched in \eqref{eq:SIS} is stable whatever $\Delta t$. However, it is costly as it requires the inversion of the operator $1+\Delta t\lt(\dxnum-c_n\pxnum[c_n]\rt)$. To summarize, among all the schemes we have investigated, we deem the PCS the most robust and least expensive one. Second, we have deliberately chosen to implement the operator $|\partial_x|$ in continuous Fourier form. It would be equally possible to discretize the integral representation of the operator, e.g., in the convenient form \eqref{eq:hilb2}. However, this may require adapted integration rules \cite{Movchan}. In this respect, a method taking advantage of FTs proves more straightforward. It also proves more versatile because the analytical form of the kernel under consideration might be unavailable in cases involving a different integrodifferential operator. The main constraint for diagonalization by the FT is that the operator be translation-invariant.

As a perspective, although definite conclusions about the validity of the approach for more general equations are yet to be obtained, we have all reasons to believe that this method applies as well to equations of the type \cite{Gui}
\begin{equation}\label{dxaplha}
\lt\{
\begin{aligned}
&-|\partial_x|^\alpha \eta(x) + \cW \partial_x \eta(x)= F'(\eta(x))   && \text{ for } x\in \mathbb{R},\\
&\eta(-\infty)=\eta_{\rm l} \quad \text{and}\quad \eta(+\infty)=\eta_{\rm r},
\end{aligned}
\right.
\end{equation}
where $F$ is a bistable nonlinearity, $\alpha \in (0,2]$, and $|\partial_x|^\alpha$ is the operator of Fourier symbol $|k|^\alpha$. However, we suspect that the preconditioning we use is insufficient for ensuring unconditional stability if $\alpha$ is larger than some value $\alpha_0 \in (1,2)$. In fact, we have explored the classical advection-reaction-diffusion case (where $\alpha=2$, and $|\partial_x|^2=-\Delta$, see Appendix \ref{Laplace}), with
\begin{equation}
\label{Flap}
F'_{\cW}(\eta)=\lt(-\cW-2 \eta\rt)\lt(1-\eta^2\rt),\qquad (|\cW|<2),
\end{equation}
where $\cW$ is the same as in the left-hand side of Equation \eqref{dxaplha}. In this case, \eqref{dxaplha} admits the analytical solution $\eta(x)=-\tanh(x)$ \cite[p.\ 291]{Griffiths}. Employing a variant of the above-described method, we have recovered a numerical approximation of this analytical solution. Our method presumably applies as well to the modified Weertman equation with gradient term \cite{Rosakis}, in which the operator $|\partial_x|$ in \eqref{Wr} is replaced by $|\partial_x|-\lambda \Delta$, where $\lambda>0$.

\acknowledgments
M.\ Josien thanks for its hospitality the CEA-DAM \^{I}le-de-France where part of this work was carried out.


\appendix
\section{Case of the Laplacian}\label{Laplace}
We briefly justify here that, with little modification, the PCS presented here to address \eqref{Wd2} can also be used for the classical reaction-advection-diffusion equation
\begin{equation}
\label{WDelta}
\lt\{
\begin{aligned}
&\partial_{xx}\eta(x)+ \cW\,\partial_x \eta(x) = F'(\eta(x))
      \quad\text{for}\quad x\in\mathbb{R},\\
      &\eta(-\infty)= \eta_{\rm l}\quad\text{and}\quad\eta(+\infty)=\eta_{\rm r},
    \end{aligned}
\right.
\end{equation}
in which the operator $-\partial_{xx}$ replaces $\dx$. Then, the counterpart of \eqref{Defdxnum} reads
  \begin{align}
    \label{dxxnum}
    -\lt(D_{xx}\mathbf{u}\rt)_j = -\partial_{xx}\etaref(x_j)+\lt(\projec \Fd^{-1} \lt\{ \kappa_p \Fd \lt\{ \injec \mathbf{v} \rt\}(k_p)  \rt\}\rt)_j,
  \end{align}
with
\begin{align}
\label{defkappa}
\kappa_p:=|k_p|^2.
\end{align}
Thus, in a first approach, one would only need to replace in \eqref{Precond2} the operator $\dxnum$ by $-D_{xx}$. The preconditioning is adapted as follows:
\begin{align}
\label{eq:PrecondLap}
M(\Delta t)\mathbf{u}=\mathcal{P}\Fd^{-1}\lt\{ \frac{\lt(\Fd\lt\{\mathcal{I} \mathbf{u}\rt\}\rt)_p}{1+\Delta t \kappa_p -\Delta t \lt(\Fd\lt\{c_n^+ \mathbf{D}_++c_n^- \mathbf{D}^-  \rt\}\rt)_p }\rt\}.
\end{align}
This modified scheme solves a dynamical equation associated with \eqref{WDelta}, similar in spirit to   (we do not write it down for conciseness). However, it turns out that this preconditioning does not ensure unconditional stability (namely, if $h$ is small, the method only converges for $\Delta t$ small). This instability is due to the truncation of the varying part $v$ of $u$ outside the box $[-L,L]$, which induces a numerical singularity in the discretization of the Laplacian near boundaries. In contrast, the preconditioning associated with $\dx$ in Section \ref{SecPCS} suffices to damp potential oscillations, probably because the latter operator is not as stiff as the Laplacian. A simple albeit costly way to overcome this stability issue is to take $\Delta t$ small.

A cheaper way is to use the classical centered discretization of the Laplacian and set
\begin{align}
\label{dxxnum2}
-\lt(D_{xx} \mathbf{u}\rt)_j =
\lt\{
\begin{aligned}
&-\partial_{xx}\etaref(x_{-m})+\frac{v_{-m}-v_{-m+1}}{h^2} && \text{if}\quad j=-m,\\
&-\partial_{xx}\etaref(x_j)+\frac{-v_{j+1}+2v_j-v_{j-1}}{h^2} && \text{if}\quad j \in \{-m+1,\cdots,m-2\},\\
&-\partial_{xx}\etaref(x_{m-1})+\frac{v_{m-1}-v_{m-2}}{h^2} && \text{if}\quad j=m-1,\\
\end{aligned}
\rt.
\end{align}
instead of \eqref{dxxnum}. The associated $\kappa_p$ that replaces \eqref{defkappa} in \eqref{eq:PrecondLap} reads
\begin{align}
\kappa_p:=h^{-2}\lt(\Fd\lt\{\lt(2,-1,0,0,\cdots,0,0,-1\rt)\rt\}\rt)_p.
\end{align}
In \eqref{dxxnum2}, the discretization at boundary points, e.g., at $j=-m$, can be explained by the following heuristic approximation:
\begin{align}
\label{BestChoice}
v''(-L)\simeq \frac{v'(-L+h/2)-v'(-L-h/2)}{h} \simeq \frac{v(-L+h)-v(L)}{h^2} =\frac{v_{-m+1}-v_{-m}}{h^2},
\end{align}
since $v(x)$ is supposed to vanish for $x<-L$. Actually, if we impose $\etaref=0$, $c_n=0$ and $F=0$, a numerical study of the eigenvalues of the PCS induced by \eqref{BestChoice} indicates that the moduli of these eigenvalues lie between $0$ and $1$ (although they can be very close to $1$), which implies that the scheme is stable. Then, with $F$ as in \eqref{Flap}, the PCS delivers outputs close to the exact solution (results not shown), although no systematic error assessment has been made in this case.
\end{document}